\documentclass[preprint,12pt,round,authoryear]{elsarticle}

\usepackage{amssymb}
\usepackage{amsmath}
\usepackage{xcolor}
\usepackage{booktabs}
\usepackage{graphicx}
\usepackage{comment}
\usepackage{multirow}
\journal{Journal of Systems and Software}

\newcommand{\Soroush}[1]{{#1}}
\newcommand{\SoroushN}[1]{{#1}}
\newcommand{\Revision}[1]{}
\newcommand{\Rem}[1]{}

\begin{document}

\begin{frontmatter}

%% Title, authors and addresses

%% use the tnoteref command within \title for footnotes;
%% use the tnotetext command for theassociated footnote;
%% use the fnref command within \author or \affiliation for footnotes;
%% use the fntext command for theassociated footnote;
%% use the corref command within \author for corresponding author footnotes;
%% use the cortext command for theassociated footnote;
%% use the ead command for the email address,
%% and the form \ead[url] for the home page:
%% \title{Title\tnoteref{label1}}
%% \tnotetext[label1]{}
%% \author{Name\corref{cor1}\fnref{label2}}
%% \ead{email address}
%% \ead[url]{home page}
%% \fntext[label2]{}
%% \cortext[cor1]{}
%% \affiliation{organization={},
%%             addressline={},
%%             city={},
%%             postcode={},
%%             state={},
%%             country={}}
%% \fntext[label3]{}

\title{CREST: Improving Interpretability and Effectiveness of Troubleshooting at Ericsson through Criterion-Specific Trouble Report Retrieval} %% Article title

%% use optional labels to link authors explicitly to addresses:
%% \author[label1,label2]{}
%% \affiliation[label1]{organization={},
%%             addressline={},
%%             city={},
%%             postcode={},
%%             state={},
%%             country={}}
%%
%% \affiliation[label2]{organization={},
%%             addressline={},
%%             city={},
%%             postcode={},
%%             state={},
%%             country={}}

\author{Soroush Javdan}
\affiliation{organization={Carleton University,},
            email={soroushjavdan@cmail.carleton.ca}, 
            city={Ottawa},
            country={Canada}}
\affiliation{organization={Ericsson Canada Inc.},
            email={soroush.javdan@ericsson.com}, 
            city={Ottawa},
            country={Canada}}

\author{Pragash Krishnamoorthy}
\affiliation{organization={Ericsson Canada Inc.},
            email={pragash.krishnamoorthy@ericsson.com}, 
            city={Ottawa},
            country={Canada}}

\author{Olga Baysal}
\affiliation{organization={Carleton University},
            email={olga.baysal@carleton.ca}, 
            city={Ottawa},
            country={Canada}}

% \affiliation{%
%   \institution{Carleton University}
%   \city{Ottawa}
%   \country{Canada}}
% \email{soroushjavdan@cmail.carleton.ca}

% \author{Pragash Krishnamoorthy}
% \affiliation{%
%   \institution{Ericsson Canada Inc.}
%   \city{Ottawa}
%   \country{Canada}}
% \email{pragash.krishnamoorthy@ericsson.com}

% \author{Olga Baysal}
% \affiliation{%
%  \institution{Carleton University}
%  \city{Ottawa}
%  \country{Canda}}
% \email{olga.baysal@carleton.ca}

%% Abstract
\begin{abstract}

The rapid evolution of the telecommunication industry necessitates efficient troubleshooting processes to maintain network reliability, software maintainability, and service quality. Trouble Reports (TRs), which document issues in Ericsson's production system, play a critical role in facilitating the timely resolution of software faults. \Rem{However, the complexity and volume of textual TR data present challenges for retrieval models, often leading to information loss due to the input size limitation of Large Language Models (LLMs).} However, the complexity and volume of TR data, along with the presence of diverse criteria that reflect different aspects of each fault, present challenges for retrieval systems.\Rem{Moreover, each TR includes distinct criteria that capture different aspects of the fault, impacting retrieval model performance.} Building on prior work at Ericsson, which utilized a two-stage workflow, comprising Initial Retrieval (IR) and Re-Ranking (RR) stages, this study investigates different TR observation criteria and their impact on the performance of retrieval models. \Rem{This study proposes}We propose \textbf{CREST} (\textbf{C}riteria-specific \textbf{R}etrieval via \textbf{E}nsemble of \textbf{S}pecialized \textbf{T}R models), a criterion-driven retrieval approach that leverages specialized models for different TR fields to improve both effectiveness and interpretability, thereby enabling quicker fault resolution and supporting software maintenance.\Rem{Utilizing these dedicated retrieval models, trained on specific TR criteria, the overall retrieval performance of the CREST improves.} CREST utilizes specialized models trained on specific TR criteria and aggregates their outputs to capture diverse and complementary signals. This approach leads to enhanced retrieval accuracy, better calibration of predicted scores, and improved interpretability by providing relevance scores for each criterion, helping users understand why specific TRs were retrieved. Using a subset of Ericsson's internal TRs, this research demonstrates that criterion-specific models significantly outperform a single model approach across key evaluation metrics. This highlights the importance of all targeted criteria used in this study for optimizing the performance of retrieval systems.
\end{abstract}

%%Graphical abstract
% \begin{graphicalabstract}
% %\includegraphics{grabs}
% \end{graphicalabstract}

%%Research highlights
% \begin{highlights}

% \item The paper proposes CREST, an ensemble of expert models trained on distinct TR criteria for enhanced retrieval.
% \item The model computes detailed relevance scores per criterion, aligned with the final aggregated score.
% \item The empirical study investigates the performance of the criterion-specific model and compares it against single-model baselines.
% \item The results demonstrate that CREST enhances both retrieval performance and confidence calibration in TR recommendations.
%\item Our approach facilitates more interpretable and transparent TR retrieval decisions in industrial settings.

%\end{highlights}

%% Keywords
\begin{keyword}
%% keywords here, in the form: keyword \sep keyword

%% PACS codes here, in the form: \PACS code \sep code

%% MSC codes here, in the form: \MSC code \sep code
%% or \MSC[2008] code \sep code (2000 is the default)

Trouble report \sep software maintenance \sep bug reports \sep information retrieval \sep neural ranking \sep natural language processing \sep telecommunications

\end{keyword}

\end{frontmatter}

%% Add \usepackage{lineno} before \begin{document} and uncomment 
%% following line to enable line numbers
%% \linenumbers
%% main text
%%

\section{Introduction}

The modern telecommunication industry is a dynamic, rapidly evolving field that relies heavily on efficient troubleshooting processes to preserve network reliability while ensuring customers receive high-quality service. 

At Ericsson, a trouble report (TR) is a critical tool for tracking information regarding the detection, characteristics, and eventual resolution of problems. TRs document issues and incidents that arise during the development and maintenance phases of the software products at Ericsson.

\Soroush{Each TR consists of multiple sections that define its characteristics. A TR typically records a \emph{headline} which serves as a short summary of the trouble, a \emph{priority} tag, the \emph{responsible team}, the \emph{product under test}, and the \emph{fault category} which indicate the defect’s type to aid TR triage, among others. However,} the most crucial section is \emph{the observation section}, which serves as a detailed explanation of the fault written by the creator of the TR. This section provides essential information for the designated team to promptly resolve the issue. The observation section is a free-text format, allowing the reporter to provide a comprehensive description of the problem. The \textbf{TR observation} includes detailed information, called criteria, describing the fault by providing all the necessary information for the fast resolution of the fault. Each of these criteria focuses on a different aspect of the fault. Common criteria include \emph{general description} of the fault, \emph{conditions} under which the fault occurred, its \emph{impact} on the system, the \emph{frequency} of the fault and Steps to \emph{reproduce} the fault. \Soroush{These criteria are considered the most crucial based on feedback from domain experts. According to their input, these criteria most often anchor discussions between reporters and responsible teams during triage and resolution. They also appear with high frequency in recent TR templates.}

To ensure the quality of TRs, they must follow a standard template and meet defined quality standards before being published for resolution. The resolution of a fault (called the answer) is only accepted after it has been validated. Previously resolved TRs (historical data) play a critical role as the information they contain can be used by quality assurance (QA) testers to improve the quality of new TRs. By quickly identifying similar past TRs and providing relevant information to designated teams, historical data enables efficient troubleshooting. Moreover, the design team can use these relevant resolved TRs during the problem analysis and correction phases, which can further reduce the time spent on debugging and testing. As a result, implementing a high-performance TR retrieval system to extract similar, previously resolved TRs can speed up software development and maintenance at Ericsson.

In recent years, there has been a notable increase in the efforts of applying machine learning (ML) and natural language processing (NLP) methods across various domains, including the telecommunication industry. Large language models (LLMs) such as bidirectional encoder representations from transformers (BERT) \citep{devlin2019bert} have shown great potential in leveraging textual data to address various tasks. Ericsson researchers proposed BERTicsson \citep{b2} as a BERT-based TR recommender system. It combines two stages: initial retrieval (IR) and re-ranking (RR). BERTicsson also uses the Sentence-BERT \citep{b3} and monoBERT \citep{b4} architecture to extract and rank a subset of historical TRs based on their relevance to new TRs. Prior work by \citet{b5} uses a similar architecture to extract and identify duplicate TRs. 
\Soroush{Prior works have also combined the headline with the observation as inputs. However, the observation was treated as one unstructured block, without parsing or modeling at the level of individual criteria, which meant its internal signals were neither isolated nor weighted explicitly.} Building on the foundational work~\citep{b2, b5} that pioneered TR retrieval studies at Ericsson, we leverage an internal LLM, specifically RoBERTa, that was trained on the internal and external telecommunication data within Ericsson.

%Building on the foundational work of \cite{b2, b5} that pioneered TR retrieval studies at Ericsson, we leverage an internal LLM, specifically RoBERTa, trained on a combination of internal and external telecommunication data.

While existing models have advanced TR retrieval, the inherent complexity and diversity of TR observations suggest there is still room for enhancement. Each TR is composed of multiple sections, with the \emph{observation section} standing out as a particularly information-rich component. This section encompasses various criteria, such as the \emph{trouble description}, \emph{impact on system}, and \emph{other criteria}, each contributing differently to the quality of retrieval. Understanding how these individual components influence retrieval outcomes is essential for identifying areas of improvement and guiding model development.
Inspired by frameworks like Branch Train Merge (BTM) \citep{b29} and DEMIX layers \citep{b30}, and DORIS-MAE~\citep{wang2023scientific}, we introduce the Criteria-specific Retrieval via Ensemble of Specialized TR models (CREST). Similar to how DORIS-MAE demonstrates the benefit of decomposing complex queries into aspects, and how BTM and DEMIX leverage expert models for modular learning, CREST trains separate models for each TR criterion. These specialized models are then combined through a weighted ensemble to improve retrieval performance.

Beyond aiming to improve retrieval performance, CREST also tackles one of the key limitations of many existing methods, which is the lack of transparency in how recommendations are generated. CREST tackles this issue by calculating separate relevance scores for each TR criterion and then combining them to determine the final ranking. This approach is designed to support interpretability \footnote{In this paper, we use the terms interpretability, explainability, and transparency interchangeably to refer to the extent to which a model’s retrieval behavior can be understood and justified by human users.} by making the influence of each TR component explicit, enabling users to better understand how recommendations are formed. In essence, CREST is designed with interpretability at its core, making the contribution of each criterion both clear and measurable.

To further explore CREST's interpretability, we investigate how well its generated detailed relevance scores align with true relevance judgments by analyzing confidence calibration. Poorly calibrated scores can undermine trust in retrieval systems, even when accuracy is high. Therefore, we assess the degree to which CREST's predicted relevance scores accurately reflect the likelihood of a correct retrieval. This calibration analysis is especially important in real-world practice, such as Ericsson, where confidence scores can guide decisions. 

\Soroush{Finally, to ensure that improvements extend beyond offline evaluation, we complement our experiments with a pilot user study at Ericsson. The study examines whether CREST's criterion-wise scores enhance the transparency of the rankings and whether the recommended TRs are perceived as credible and practically useful during triage, rather than merely appearing more accurate in offline metrics.

In summary, our analysis shows that observation criteria contribute unevenly across different retrieval stages. Building on this, our criterion-specific ensemble (CREST) consistently outperforms single-model approaches in the two-stage workflow and remains superior even when evaluated in isolation. Moreover, CREST improves confidence calibration and provides criterion-wise relevance scores, making recommendations both more reliable and more interpretable for engineering decision-making.}

Using Ericsson's internal trouble-reporting dataset, we answer the following research questions:

\begin{enumerate}

\item RQ1: \emph{What impact does each TR criterion have on retrieval model performance}? 

\item RQ2: \emph{To what extent does CREST improve the retrieval performance compared to a single model}?

\item RQ3: \emph{How does the calibration of relevance scores produced by CREST compare to those generated by the criterion-agnostic model}?

\Soroush{\item RQ4: \emph{What are the users' perceptions and rankings of CREST in terms of transparency, usefulness, credibility, and accuracy}?}

\end{enumerate}

The remainder of this paper is organized as follows: Section~\ref{sec:background} highlights background relevant to this study and summarizes related work. Section~\ref{sec:method} presents the details of the study approach and includes an overview of retrieval systems, data related to TRs, the initial retrieval (IR) stage, the re-ranking (RR) stage, training, and inference. Section~\ref{sec:eval} presents the evaluation procedure, including evaluation strategy, metrics, and datasets. Section~\ref{sec:results} reports the results of our empirical study. Finally, Section~\ref{sec:conclusion} concludes the paper by discussing possible future research directions.

\section{Background and Related Work} \label{sec:background}

Since we leverage Large Language Models (LLMs) in this work, this section provides an overview of LLMs, highlighting their strengths and limitations, as well as their application within the telecommunications industry. Additionally, it discusses recent research on the use of LLM-based models for text retrieval problems.

\subsection{Large Language Models}

\Revision{Natural Language Processing (NLP) has transformed with the rise of LLMs, which excel in complex tasks like text classification, summarization, and semantic understanding. Their capabilities make them well-suited for processing unstructured textual data in TRs, including in telecommunications. Among the most prominent LLMs are BERT (Bidirectional Encoder Representations from Transformers) \citep{b1}, RoBERTa (a Robustly optimized BERT approach) \citep{b6}, T5 (text-to-text Transfer Transformer) \citep{b7}, and ERNIE (Enhanced Language Representation with Informative Entities) \citep{b8}.These models are well capable of recognizing complex language patterns, which makes them useful for tasks such as analyzing and categorizing Trouble Reports (TRs). Their ability to process and understand extensive and complex datasets can significantly streamline operations in telecommunications, reducing the time required for manual data handling~\citep{b28}.
}

\Soroush{Natural Language Processing (NLP) has been reshaped by the rise of large language models (LLMs), which excel at tasks such as text classification, summarization, and semantic understanding. Cutting-edge models like the GPT series \citep{achiam2023gpt}, LLaMA3 \citep{grattafiori2024llama}, DeepSeek \citep{liu2024deepseek}, Mistral \citep{jiang2023mistral7b}, and Qwen \citep{qwen2.5} are now increasingly integrated into retrieval pipelines, serving roles as encoders, re-rankers, or generative reasoners. In parallel, classic compact transformers including BERT \citep{b1}, RoBERTa \citep{b6}, T5 \citep{b7}, and ERNIE \citep{b8} remain widely deployed in production due to strong efficiency–accuracy trade-offs and mature tooling~\citep{puthenputhussery2025large}.

In this study, we focus on Ericsson's domain-trained RoBERTa (TeleRoBERTa) for two main reasons. TeleRoBERTa, trained on proprietary telecom corpora, matches the performance of much larger foundation LLMs on telecom standards QA benchmarks while using an order of magnitude fewer parameters. This makes it particularly well aligned with the distributions found in telecom-specific tasks \citep{karapantelakis2024using}. In addition, our objective is criterion-aware retrieval that is both efficient and interpretable, and that can be immediately integrated into Ericsson's existing two-stage pipeline. Fine-tuning a compact, domain-adapted encoder not only reduces compute and latency overhead but also enables drop‑in improvements within the existing production stack.
}

\subsection{Neural Retrieval Models}

\Soroush{Neural retrieval models (NRMs) for document search largely fall into two families: \emph{cross-encoders} and \emph{bi-encoders}. Cross-encoders (e.g., monoBERT~\citep{b4}) score a query–document pair jointly, yielding strong relevance estimates but high computational cost; multi-stage pipelines therefore retrieve with BM25 and re-rank with monoBERT or duoBERT’s pairwise comparator~\citep{b9}. To reduce cost, bi-encoders encode queries and documents independently (e.g., TwinBERT~\citep{b10} with a Siamese setup~\citep{b11}), enabling offline document embeddings for fast initial retrieval; late-interaction models like ColBERT~\citep{b12} bridge the two by combining separate encodings with token-level matching.
Within Ericsson, BERTicsson~\citep{b2} applied this multi-stage recipe to recommend solutions for new TRs, outperforming BM25~\citep{b13}. Related work on duplicate TR detection~\citep{b5} explored domain adaptation~\citep{b14}, sequence learning~\citep{b14}, multi-stage fine-tuning, and elastic weight consolidation~\citep{b15} to mitigate catastrophic forgetting.

All prior systems treat a TR observation as a single block of text. This monolithic view obscures criterion-specific signals and limits interpretability of relevance scores. These gaps motivate our approach which explicitly models observation criteria and aggregates their signals to improve both retrieval quality and transparency.}

\subsection{Transparency and Explainability in Neural Retrieval Models}

While neural retrieval models have achieved impressive performance across a variety of information retrieval tasks, they often operate as black boxes, offering limited insights into how specific relevance decisions are made \citep{rudin2019stop}. This lack of transparency poses significant challenges in domains such as software maintenance, where practitioners require justifiable and interpretable recommendations to support critical decision-making.

Broadly, approaches to explainability in neural retrieval can be grouped into two categories: post-hoc and design-time strategies \citep{anand2022explainable}. Post-hoc methods attempt to interpret trained models by analyzing model behaviour or internal representations after training is complete. These include techniques such as feature attribution~\citep{wang2024quids, zhang2020query}, attention mechanism~\citep{lucchese2023can}, and representation probing~\citep{wallat2023probing}, each aimed at uncovering which input features or structures most influence the model's outputs. While insightful, these approaches often operate heuristically and may not yield stable or actionable explanations across instances or domains.

On the other hand, models that explicitly integrate explainability into their architecture offer a more systematic path toward interpretability. These models often utilize modular structures or enforce architectural constraints to ensure that each part of the decision process can be traced and understood. In the context of neural retrieval, such approaches may involve segmenting the relevance estimation process into interpretable components aligned with domain-specific dimensions of the input \citep{yu2022explainable,leonhardt2023extractive}.

In the context of TR retrieval, we adopt a similar strategy through the proposed CREST model, which generates criterion-specific relevance scores that are combined into a final ranking. This design improves both retrieval accuracy and interpretability, allowing users to trace each recommendation back to distinct TR criteria. In this way, CREST promotes explainability by design, ensuring that each component's contribution to the final output is both meaningful and measurable.

\subsection{Confidence Calibration of Neural Ranking Models}

Neural rankers often output relevance scores that are not well-aligned with the actual likelihood of relevance \citep{penha-hauff-2021-calibration}. While these scores can be effective for sorting documents, the lack of calibration can limit their interpretability and downstream utility, especially in settings where scores are aggregated, thresholded, or presented to users. Calibration analysis allows us to assess how well the predicted scores reflect true probabilities, providing a clearer picture of a model's reliability.

In retrieval systems that rely on multiple independent signals, such as those derived from different aspects or criteria, calibrated outputs are particularly valuable. When models produce confidence scores that are better aligned with observed relevance, combining their outputs becomes more meaningful and consistent. Calibrated scores also support more robust decision-making when setting thresholds or ranking candidates across criteria. Thus, even in systems that do not explicitly optimize for calibration, understanding and measuring this property is key to ensuring the reliability of the produced relevance scores.

\subsection{Related Work} \label{sec:rw}

The text retrieval task has seen substantial advancements with the advent of large language models (LLMs). Recently, researchers have applied different techniques to improve LLM-based text retrieval systems. RocketQAv2 \citep{b20} and AR2 \citep{b21} use a joint training strategy for both the retriever and re-ranker parts. RocketQAv2 jointly trains the retriever and the re-ranker with a focus on reducing the difference between their relevance prediction distributions. AR2 formulates the text-ranking problem using a Generative Adversarial Network (GAN)~\citep{b22}. It jointly optimizes a retriever as the generative model and a ranker as the discriminative model, with the retriever generating hard negatives to improve ranking performance. Poly-encoder~\citep{b23}, HLATR~\citep{b24}, and Uni-encoder~\citep{b25} aim to balance the efficiency of bi-encoders with the rich interaction capabilities of cross-encoders. RankLLaMA~\citep{b27} was introduced as a fine-tuned LLaMA model for multi-stage text retrieval, demonstrating that more advanced LLMs can outperform smaller models. DORIS-MAE~\citep{wang2023scientific} tackles the limitations of NIR models trained on simple queries, which often fail on complex, multifaceted inputs. It introduced a benchmark dataset with hierarchical aspect-based queries in the scientific domain and demonstrated that aspect decomposition can improve retrieval performance.
Parameter-efficient fine-tuning (PEFT) techniques have also gained traction in text retrieval. \citet{b26} employed prompt tuning techniques, demonstrating that by updating only a small portion of the model parameters, performance comparable to conventional fine-tuning can be achieved. This method significantly enhances out-of-domain generalization and improves confidence calibration. \citet{b19} introduced a Semi-Siamese Bi-encoder Neural Ranking Model utilizing PEFT techniques, which has shown significant improvement by updating only a small portion of the model parameters.

In the realm of transparency and explainability, several approaches have been proposed to make neural retrieval models more interpretable. \citet{leonhardt2023extractive} introduced the Select-and-Rank paradigm, wherein the model selects a subset of sentences from a document as an explanation and then uses this selection exclusively for prediction. This approach treats explanations as integral components of the ranking process, enhancing interpretability by design. \citet{zhang2021explain} proposed ExPred, a model employing multi-task learning during the explanation generation phase. It balances explanation and prediction losses and subsequently utilizes a separate prediction network trained solely on the extracted explanations to optimize task performance. \citet{wallat2023probing} conducted an in-depth analysis of BERT's ranking capabilities by probing its internal representations. Their study reveals insights into BERT's effectiveness in ranking tasks and highlights areas for improvement in aligning its internal mechanisms with established information retrieval principles.

Recent studies have also emphasized the importance of calibration in neural rankers. \citet{penha2021calibration} investigated the calibration and uncertainty of neural retrieval models in conversational search, highlighting that neural rankers often produce overconfident predictions. \citet{tam2023parameter} proposed prompt tuning methods that improve generalization and calibration of dense retrievers with minimal parameter overhead. \citet{yu2024explain} used LLMs to generate natural language explanations and applied Monte Carlo sampling to achieve better scale calibration, while maintaining or improving ranking performance.

These advancements highlight the continuous evolution and optimization of text retrieval systems using LLMs and related techniques. \Soroush{CREST takes a different approach compared to earlier interpretable or LLM-based retrieval systems like ExPred~\citep{zhang2021explain} and RankLLaMA~\citep{b27}, which mainly aim to improve explanation generation or scale single end-to-end rankers. Instead, CREST introduces a modular, multi-aspect retrieval framework. Rather than depending on one model to capture all query semantics, it breaks down each query into predefined criteria and combines the outputs of specialized models trained for those aspects. This structure enables CREST to deliver stronger retrieval performance and clearer interpretability, since each result's relevance can be directly tied to specific criteria.}

\section{Methodology} \label{sec:method}

In this study, we conduct a comprehensive investigation to understand how different Trouble Report (TR) observation criteria affect the performance of TR retrieval models such as BERTicsson~\citep{b2}.

This study also aims to train criterion-specific models and aggregate them to optimize the retrieval process, enhancing the system's ability to attend to information from different criteria. The TR observations are preprocessed and parsed using a standardized Ericsson TR template to extract various informative criteria. The impact of each criterion is assessed on the overall retrieval system's performance. \Rem{This also helps to alleviate the limitations posed by encoders input length.} Moreover, this approach provides transparency in the decision-making process by providing separate relevance scores to each observation criterion, enabling users to trace retrieval outcomes back to meaningful components of the input. %The TR observations are preprocessed and parsed using a standardized Ericsson TR template to extract various informative criteria. The impact of each criterion is assessed on the overall retrieval system's performance.

\subsection{Two-stage TR Retrieval Model}

This methodology employs a two-stage ranking architecture similar to BERTicsson where the models used are adapted with RoBERTa, resulting in TwinRoBERTa or ColRoBERTa for initial retrieval and monoRoBERTa for re-ranking stage.
The first stage utilizes a bi-encoder architecture, which is less computationally intensive and allows faster processing. This efficiency arises from the ability to compare pre-calculated document embeddings with the embeddings of new queries. The processed text from the preprocessing step serves as the input for the IR stage, which retrieves a top-K list of candidate TRs ranked by the relevance of their accepted answers to the target TR observation.

The second stage involves re-ranking (RR) the candidates provided by the IR stage. Here, each candidate from the top-K list is paired with the query and processed through the monoRoBERTa model, a cross-encoder that provides a more detailed and computationally intensive comparison. This re-ranking stage benefits from the cross-encoder's ability to assess finer details within the interactions between the query and document, resulting in a highly accurate, final ranked list of relevant TRs.

\begin{figure}[!t]
	\centerline{\includegraphics[width=0.9\columnwidth]{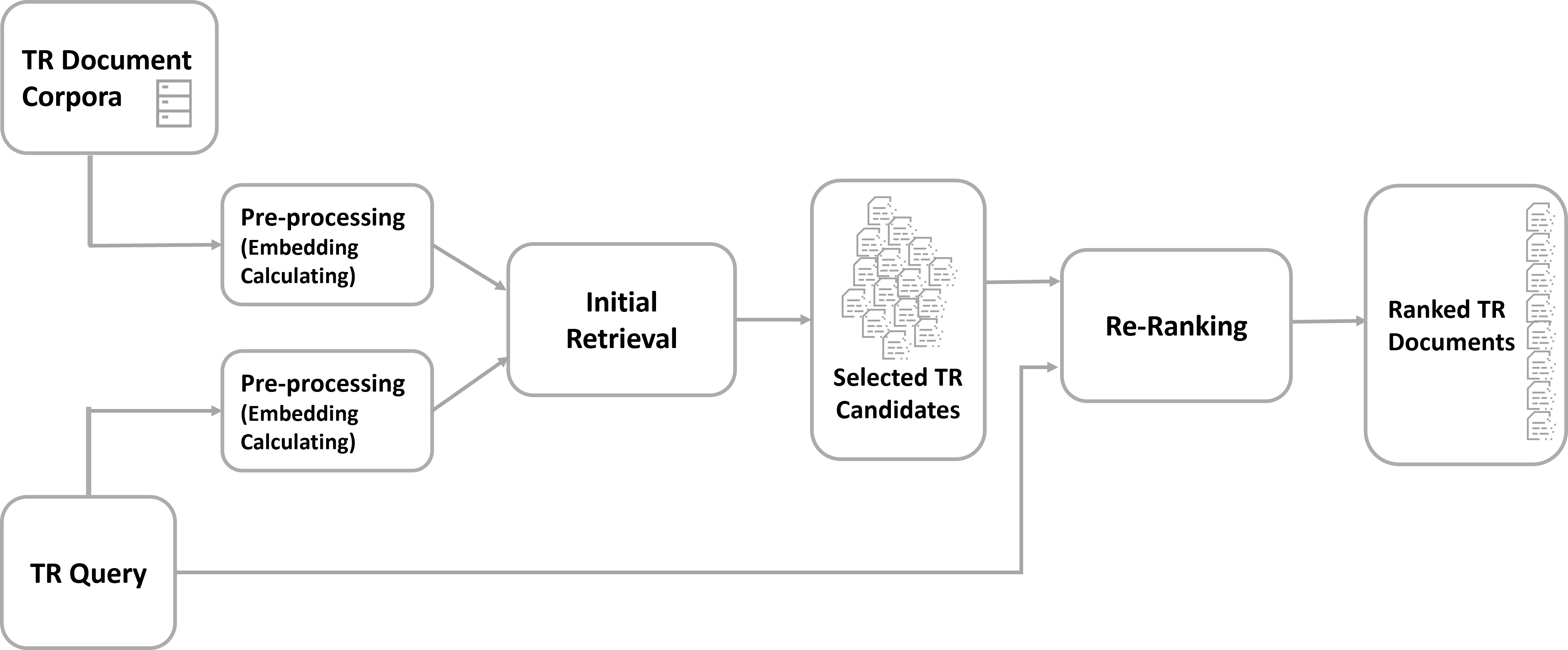}}
	\caption{Overview of the utilized TR recommendation system.}
	\label{fig_over}
\end{figure}

Figure~\ref{fig_over} provides an overview of the TR retrieval system utilized in this study, illustrating the data flow through the multi-stage architecture. The workflow begins with the preprocessing of TRs and the calculation of embeddings. Document embeddings for the corpus are computed once and used throughout the IR step. The query embeddings are similarly generated and used to select the top-K most similar TR documents. These selected documents are then paired with the query for the re-ranking step, executed by monoRoBERTa.

This two-stage strategy efficiently integrates the speed and lower computational demands of a bi-encoder in the IR phase, with the accuracy and depth of analysis provided by a cross-encoder in the RR phase.

Criteria-specific Retrieval via Ensemble of Specialized TR models (CREST) adopts a similar two-stage approach but distinguishes itself by utilizing an aggregation of TR retrieval models, each trained with a different TR observation criterion and specializing in retrieving documents based on that specific aspect. This ensemble setup enables the retrieval system to attend to various facets of the queries and retrieve documents relevant to each criterion. A document's final relevance score is then computed through a weighted aggregation of its individual criterion-specific scores, ensuring that the distinct contributions of each criterion are accurately captured and utilized.

This design not only enhances the overall effectiveness of the TR retrieval system by leveraging the strengths of specialized models but also provides individual relevance scores per criterion, offering transparency and traceability in the decision-making process for end-users.

\subsection{Trouble Report Data}

In this study, we utilize historically resolved trouble reports (TRs) as the training data for both the IR and RR models. We primarily focus on the headline and observation sections of the TRs, which can describe the problem presented and are used as queries to retrieve relevant documents. The accepted answers, which detail the resolutions to these problems, serve as the documents in our retrieval system. This setup allows us to rank previously resolved TRs based on their relevance to new queries that are formulated from the headline and observation sections of new TRs. Notably, while our approach is centered around retrieving TRs based on the relevancy of their answers to the new TRs observation, its potential applications extend beyond this scope. For instance, our methodology can be adapted for identifying TR duplicates~\citep{b5}, where the emphasis lies on finding similar TRs based on the similarity of their observations.

A typical TR at Ericsson contains the following sections:

\begin{enumerate}
	\item \emph{Headline}: A sentence summarizing the problem, often containing critical information about the issue.

	\item \emph{Observation}: Detailed text that describes the problem comprehensively. This free-format text includes vital details for the responsible team, such as the general description of the tester's observation, \textit{impact} on the system, \textit{conditions} causing the issue, \textit{frequency} of the problem, and \textit{reproducibility}.
	
	\item \emph{Answer}: Extensive text that explains the root cause and resolution.

	\item \emph{Faulty Product Detail}: Information about the fault, the affected software product, and additional relevant details. 
\end{enumerate}

\Soroush{Parsing is applied to the TR observation to extract multiple criteria and assess their contributions to retrieval. We use a lightweight Python regex parser that targets the standardized observation template, looking for headers explaining different criteria. Although the observation is free text, testers complete a consistent, organization-wide template. Fields are optional, so some entries can be missing or not be complete. Ericsson's internal quality-control review takes place before TR is being published, which limits format drift and typographical variation. In practice, the parser is reliable, with only occasional edge cases such as merged or empty fields} We select the following specific criteria for this study:

\begin{enumerate}
	\item \emph{Trouble Description}: This is the most informative part of the observation, as it explains the problem in detail.
	\item \emph{Impact}: A short text explaining the impact of a new problem on the system.
	\item \emph{Condition}: The condition in which the problem occurred.
	\item \emph{Frequency}: The frequency of observing the problem.
	\item \emph{Steps to reproduce}: Explain if the problem is reproducible and how it can be reproduced.
\end{enumerate}

\begin{figure}[!h]
\centerline{\includegraphics[height=0.6\columnwidth]{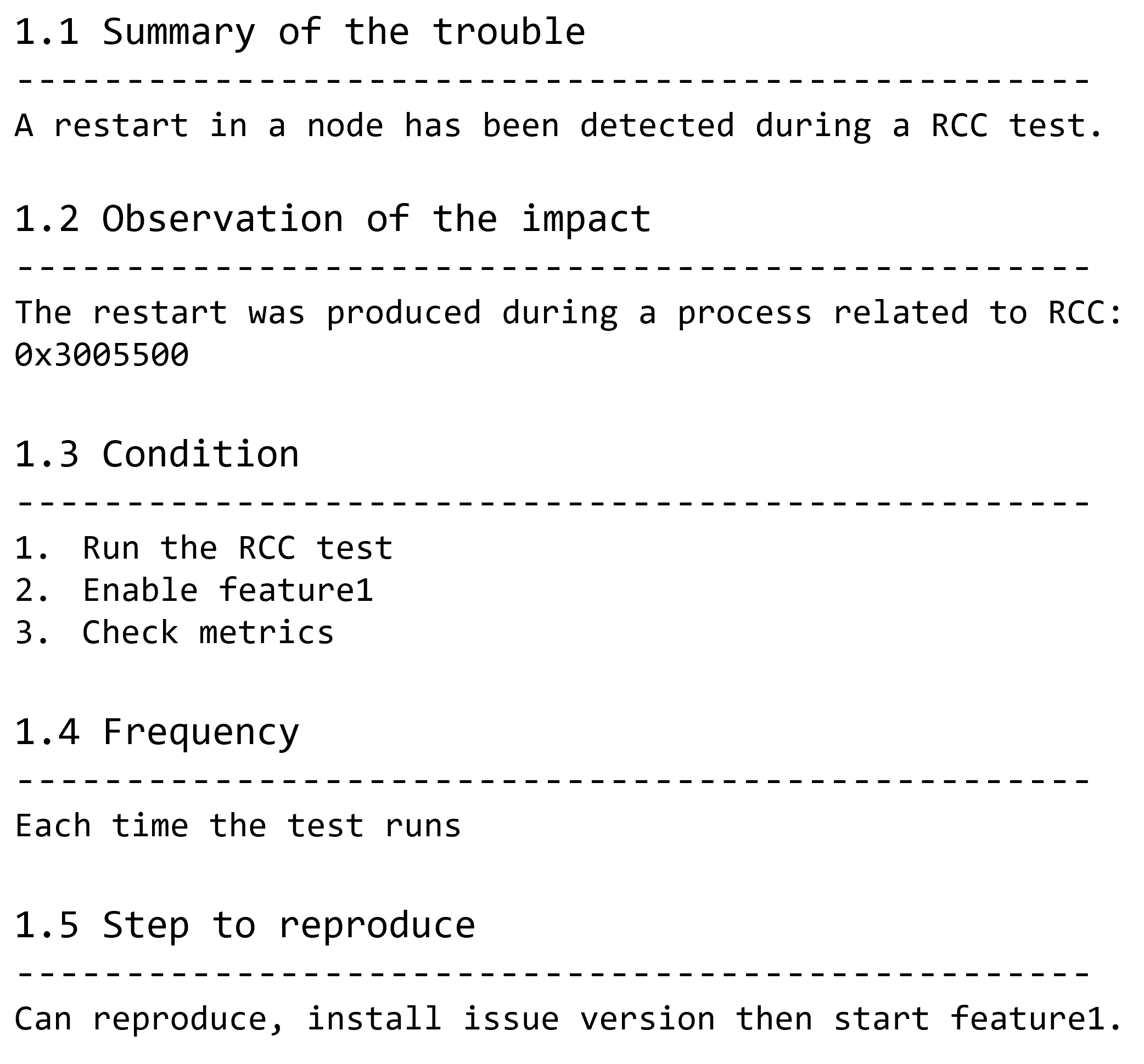}}
\caption{An example of the TR observation field with different criteria.}
	\label{tr_example}
\end{figure}

\begin{figure}[!h]
	\centerline{\includegraphics[width=0.8\columnwidth]{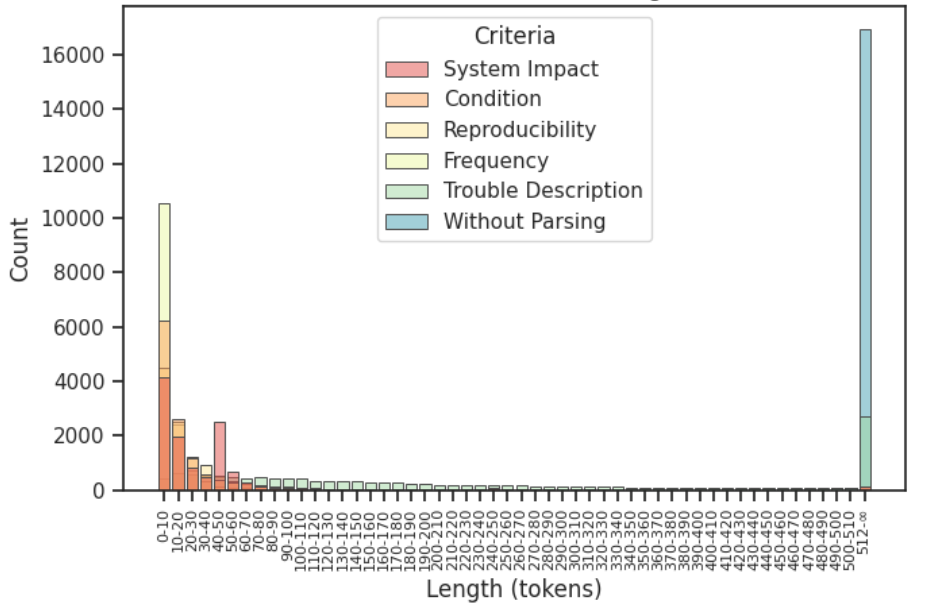}}
	\caption{Distribution of TR observation criteria based on the token length.}
	\label{fields_hist}
\end{figure}

Figure~\ref{tr_example} illustrates an example of the TR observation. Although it is of free format, the observation follows a template that assists writers in structuring and positioning its various criteria. 
Figure~\ref{fields_hist} shows the token length distribution of observation section criteria before and after parsing. While unparsed observations often contain long, unstructured text, the parsed fields are typically more concise and structured. This reinforces the importance of evaluating each field's individual contribution to retrieval effectiveness, offering a more targeted alternative to treating the observation as a monolithic input.

With this in mind, we generate multiple datasets that pair the TR headline with different parsed observation criteria. This setup allows us to investigate the retrieval value of each field and gain a better understanding of its role in the overall retrieval task. Moreover, these insights not only support the design of our criterion-specific modeling approach but also offer practical guidance: if certain criteria are shown to have a stronger impact on retrieval quality, TR creators can be encouraged to elaborate more on those fields, ultimately improving the effectiveness of the retrieval system.

\subsection{Initial Retrieval}

In the initial retrieval (IR) stage, pre-processed queries and documents extracted from TRs are utilized to generate a top-K candidate list. The initial retrieval of candidates within the top-K list is not essential, as a later re-ranking stage will adjust their order. However, it is crucial that the IR stage includes as many relevant documents as possible within the top-K list to achieve a high overall system performance.  For the retrieval of the top-K candidates, a bi-encoder architecture is employed, \Soroush{specifically the TwinRoBERTa and ColRoBERTa models. These models are frequently used in various information retrieval and similarity comparison tasks (e.g., BERTicsson \citep{b2})}. Their main objective is to determine the similarity between two inputs. In our context, these inputs are a TR observation (including headlines and various criteria) and a TR answer. They effectively separate the query and document processing by employing distinct encoders for each, followed by a mean pooling layer to generate fixed-length representations for both inputs. To further enhance domain-specific performance, TeleRoBERTa~\citep{b1} is incorporated, a version of the RoBERTa model further trained in telecommunications data. The domain-specific knowledge of this model can benefit the TR retrieval system performance \citep{nimara2024entity}.

The similarity score between query and document embeddings is calculated using a fully connected layer. The resulting score is then used to rank and select the top-K candidates. For training, the model is exposed to relevant and irrelevant query and document pairs in order to adjust its weights, ensuring higher scores for relevant pairs and lower scores for irrelevant ones.

To minimize latency during inference, the embeddings of all documents in the corpus are pre-calculated by leveraging the decoupled nature of TwinRoBERTa and ColRoBERTa, which independently process queries and documents. This allows for only computing the query representation during the inference and comparing it against all pre-stored document embeddings. This possibility makes models that follow bi-encoder architectures significantly faster than cross-encoder architectures like monoRoBERTa, in which both query and document are processed simultaneously.

%\st{5CREST is applied at both the IR and RR stages. In the IR stage, CREST has potential to improve the retrieval performance while keeping the same latency level, as the embeddings of all documents are pre-calculated for each criterion-specific model, and only the query embedding needs to be calculated for each model.}
%\Soroush{Moved to Inference section}

\subsection{Re-Ranking Stage}

In the re-ranking (RR) stage, the IR output, which is the top-K candidates, is further processed for a more precise ranking. \Soroush{In practice, we generate this top-K with the strongest IR configuration on our validation set, and use that retriever's output as input to RR, so the cross-encoder operates on the best available candidate set.} Following the approach used in BERTicsson \citep{b2}, this study uses a cross-encoder architecture, specifically adopting the monoRoBERTa framework for the RR stage. Similar to the IR stage, TeleRoBERTa is employed here to leverage its telecommunications domain knowledge.

The input of the monoRoBERTa model is a concatenated string of a query and document tokens separated by a special token as defined in large language models, e.g., ``[CLS], query tokens, [SEP], document tokens, [SEP]''. Since both the query and document can contain important contextual information, we allocate tokens equally between them to preserve balanced representation.

%In response, our study is actively exploring how various TR criteria influence the retrieval system's efficiency.

Unlike bi-encoder models, which use a decoupled approach to represent queries and documents, the cross-encoder architecture integrates the processing of query and document representations. Since cross-encoder models compute query and document representations together, pre-computing embeddings is not feasible, resulting in higher retrieval latency. However, the increased latency remains minimal, as the model only applies to the top-K candidates identified in the IR stage, which is a subset considerably smaller than the entire TR corpus.

During training, the cross-encoder receives pairs of relevant and irrelevant queries and documents, adjusting its weights to enhance its performance. In the inference phase, the model only processes receiving query pairing with the top-K candidates extracted during the IR stage. By limiting the number of query-document pairs processing,\Rem{we effectively mitigate the computational latency inherent to cross-encoder models.} the computational latency inherent to cross-encoder models is effectively mitigated.

Similar to the IR stage, applying CREST in the RR stage can improve performance. However, unlike the IR stage, using CREST in the RR stage will introduce additional latency, as the top-K candidates must be processed by each criterion-specific model. This increases latency by a factor equal to the total number of criterion-specific models. This added latency is acceptable, as the ensemble size is small, with CREST utilizing only four criterion-specific models.

\subsection{Criteria-specific Retrieval via Ensemble of Specialized TR models (CREST)}

\begin{figure}[!t]
	\centerline{\includegraphics[width=\columnwidth]{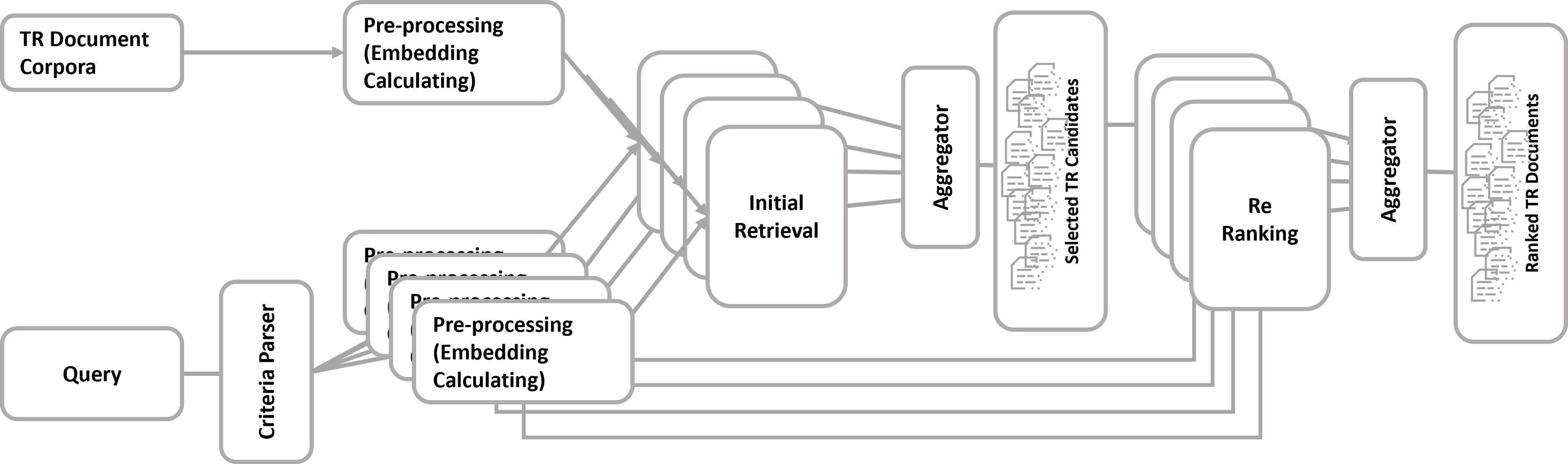}}
	\caption{\Soroush{Overview of CREST in a two-stage pipeline: bi-encoders (Twin/ColRoBERTa) retrieve top-K candidates and a cross-encoder (monoRoBERTa) re-ranks them. Unlike the baseline criteria-agnostic two-stage workflow shown in Figure~\ref{fig_over}, CREST adds criterion-specific models whose per-criterion scores are aggregated into the final relevance score.}}
	\label{ensamble}
\end{figure}

In this study, we introduce the Criteria-specific Retrieval via Ensemble of Specialized TR models (CREST), an ensemble-based framework designed not only to enhance retrieval performance but also to improve transparency (aka interpretability) in the retrieval process. By leveraging the structured nature of TR observations, each model in the ensemble is trained using a distinct criterion extracted from the observation section, enabling it to specialize in handling a specific type of information. This setup allows CREST to capture diverse aspects of the TR content and aggregate them to form a more comprehensive and interpretable relevance signal.

CREST supports both bi-encoder (TwinRoBERTa, ColRoBERta) and cross-encoder (monoRoBERTa) architectures. \SoroushN{For each criterion, a separate query is generated, and its corresponding model is used to evaluate document relevance. These individual scores are then combined through a learned, criterion-specific weighted aggregation, implemented as a linear combination of criterion-specific scores with non-negative weights in range of [0,1] to produce the final relevance score for each TR document. These aggregation weights are optimized in a separate training stage for each cross-encoder and bi-encoder models, while the models parameters are kept frozen. The optimization uses a hinge loss applied over the aggregated scores, and the final aggregation model is selected based on the Mean Reciprocal Rank (MRR) achieved on the validation set.} This mechanism ensures that the specific contribution of each criterion is reflected in the final ranking, offering greater transparency and traceability in the retrieval process. Figure~\ref{ensamble} provides an overview of the CREST setup.

% For each criterion, a separate query is generated, and its corresponding model is used to evaluate document relevance. These individual scores are then combined through a \Soroush{learned, criterion-specific weighted aggregation} to produce the final relevance score for each TR document. \Soroush{Weights are estimated on a held-out validation split to maximize ranking quality.} This mechanism ensures that the specific contribution of each criterion is reflected in the final ranking, offering greater transparency and traceability in the retrieval process. Figure~\ref{ensamble} provides an overview of the CREST setup.

\begin{figure}[!t]
	\centering
   	\includegraphics[width=0.95\columnwidth]{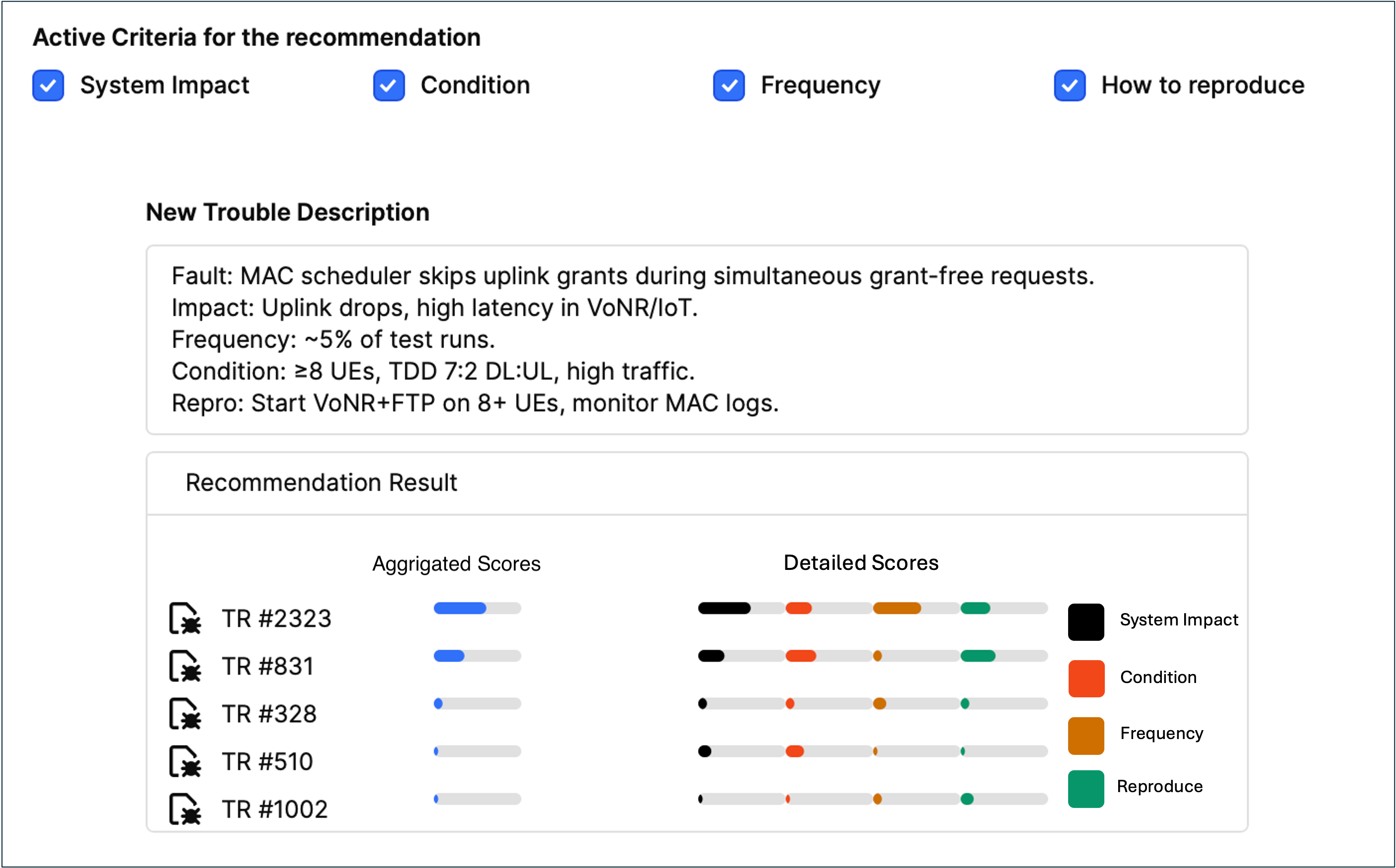}
	\caption{\Soroush{Mockup of the CREST interface showing selectable criteria and both disaggregated (per-criterion) and aggregated relevance scores, enabling configurable focus and clearer rationale for retrieved results.}}
	\label{fig:CREST-interface}
\end{figure}

Figure~\ref{fig:CREST-interface} illustrates what CREST may look like in practice. Users can interactively select which criteria to activate, such as system impact, condition, frequency, or how to reproduce, depending on the context or diagnostic goal. Once a new trouble description is entered, CREST generates criterion-specific queries, scores each candidate TR using its corresponding model, and visualizes both the aggregated and disaggregated scores. The coloured bar indicators enable users to assess the relative contribution of each criterion to the final score, providing transparency and control over the retrieval process. The framework is adaptive to missing inputs; if certain criteria are unavailable during inference, only the relevant specialized models are triggered. Likewise, users can configure the system to focus solely on a particular criterion when diagnosing a specific issue or activate all available criteria to maximize overall performance. This flexibility makes CREST suitable for a range of retrieval scenarios and user needs.

CREST is not positioned as a replacement for existing retrieval models but rather as a modular enhancement that can be integrated into various TR retrieval pipelines. Its criterion-specific decomposition and aggregation strategy make it a flexible solution that can extend to systems based on other LLMs.

\subsection{Training} \label{sec:train}

The training process of the TR retrieval system involves training both the bi-encoder and cross-encoder models in a supervised manner. For training, queries, positive documents, and negative documents are created from the extracted TRs. The query is the description of the issue mentioned in the TR (e.g., headline, and various criteria). The positive document is the answer section paired with the same TR, while the negative document is the same query from the TR, paired with the answer from a different TR. The result is two pairs:  $<\textrm{query}, \textrm{positive document}>$ (called a relevant pair) and $<\textrm{query}, \textrm{negative document}>$ (called an irrelevant pair). The final training datasets maintain a 1:1 ratio for positive and negative pairs, meaning that for each collected TR, there is one positive pair and one negative pair.

\Rem{
The goal of this study is to explore the effect of various TR observation criteria on retrieval performance and train a criterion-specific model. To achieve this, distinct datasets are created based on each criterion to train and evaluate the TR retrieval system. This approach serves two purposes: first, to measure the impact of each criterion on retrieval system performance, and second, to evaluate the performance effect of having an ensembled model, aggregating each of the criterion-specific models.} The goal of this study is to explore the effect of various TR observation criteria on retrieval performance and train criterion-specific models. To support this, we construct separate datasets for each observation criterion, allowing us to both train specialized retrieval models and assess the individual contribution of each field to the overall system. This setup enables a detailed evaluation of criterion-level impact while also serving as the foundation for the proposed ensemble framework, where models trained on different criteria are later aggregated.

In total, two types of models were trained in this study: TwinRoBERTa, as a bi-encoder, and monoRoBERTa, as a cross-encoder. All models use the same triplet hinge loss function \citep{b19}, which leverages the relevance scores calculated for both positive and negative pairs to optimize the ranking performance. We used a batch size of 64 and the Adam optimizer with a learning rate of $10^{-5}$ for all encoders.

\subsection{Inference} \label{sec:infer}

During the inference phase, the TR retrieval system begins the ranking process once the tester initiates a new TR with both headline and observation details. The query is formed by combining the headline and available criteria from the observation section. 

The IR stage starts by computing the query that is then used to calculate relevance scores for different documents, extracting the top-K candidates. During the RR stage, the system re-orders the extracted candidates from the IR stage. Each candidate is paired with the query, and the pair is processed through a cross-encoder model to compute a new relevance score, which is used to refine the ranking of top-K candidates. 

Adjusting the value of K affects both performance and latency. A higher K increases the likelihood of capturing relevant documents but also increases the input for the RR stage, thus increasing latency.\Rem{ Consequently, the value of K should be considered a critical hyperparameter.}

For both IR and RR stages, the number of criterion-specific models used in CREST is determined by the availability of the observation criteria after parsing the observation entered by the tester. If the TR observation contains all criteria, then all models in CREST are active. This helps to ensure that the system uses all available information to optimize retrieval performance.

\Revision{As the document embeddings are pre-calculated for the IR stage, only the query embedding needs to be computed for all of CREST's criterion-specific models. Consequently, CREST maintains the same latency level as individual models. In contrast, for RR, CREST multiplies the latency by the total number of criterion-specific models. However, by running the models in parallel on two GPUs, we effectively reduce this latency by half, ensuring the system remains efficient. Table \ref{CREST_computation_costs} summarizes the computation and latency costs for both training and inference stages.}

\Soroush{As the document embeddings are pre-calculated for the IR stage, only query embeddings need to be generated at runtime. In CREST, this involves four query embeddings instead of one, but they are computed in parallel so the latency remains close to that of a single model. At a large scale, this keeps first-stage retrieval feasible on very large TR corpora, since each query’s four embeddings are matched against pre-computed indexes to select a small candidate set for reranking (RR).  

The RR stage is more expensive because CREST scores every candidate with all criterion-specific models. We manage this by keeping the candidate set relatively small and by batching candidates so each model processes them in one pass. Running models across two GPUs further reduces wall-clock time and keeps end-to-end latency within acceptable service windows for production. The number of criteria can also be reduced for tasks that require very low latency, which creates a trade-off between latency and performance. Table~\ref{CREST_computation_costs} presents training and inference costs.}

\begin{table}[t!]
\centering
\begin{tabular}{lcc}
\toprule
Stage         & Single Model       & CREST                  \\
\midrule
IR and RR(Training)      & Single training             & 4 separate trainings            \\
IR(Inference) & 1 LLM pass                     & 4 LLM passes                        \\
RR(Inference) & $k$ LLM passes                 & $4*k$ LLM passes                     \\
\bottomrule
\end{tabular}
\caption{Comparison of computation and latency costs between the single model and the CREST model, with \textbf{k} representing the number of candidates returned in the IR stage.}
\label{CREST_computation_costs}
\end{table}

As presented in Table \ref{CREST_computation_costs}, the training latency for CREST is four times higher than that of a single model when using a single GPU, due to the need to train four separate criterion-specific models. However, we leveraged four GPUs to run these trainings concurrently, which kept the overall training time comparable to that of a single model, while increasing GPU resource usage accordingly.

\section{Evaluation} \label{sec:eval}

In this study, we create distinct datasets incorporating diverse information for both training and evaluation. We also determine the impact of various TR observation criteria on retrieval system performance and the enhancement that the CREST can bring to the retrieval system.

\subsection{Datasets}\label{sec:dataset}

To construct datasets, trouble reports were organized into groups based on the presence of specific criteria within their observation sections, as illustrated in Figure~\ref{tr_example}. Due to the heterogeneous nature of TRs, not all contain identical observation fields, which affects the dataset's composition. We consistently include the headline section and trouble description criteria in all datasets to preserve essential information. \Soroush{Approximately $60\%$ of the parsed TRs used for this study contain all listed criteria, and this coverage is significantly higher among more recent TRs than among older reports partially included in this study.}

To prepare data for the criterion-specific models and analyze the significance of each of the individual criteria, the following approach was applied for each criterion:

\begin{enumerate}
	\item Filtering TRs to isolate those containing the criterion under study.
	\item Forming queries and documents from this refined subset to produce datasets focused on that specific criterion. Documents are pre-processed versions of TR answers.
	\item Combining the ``headline'' and ``trouble description" with the criteria under evaluation to formulate queries, ensuring each dataset is tailored to our research focus.
\end{enumerate}

A baseline dataset is created for each experimental set, consisting of TRs with queries derived solely from the ``headline'' and ``trouble description''. This baseline allows for direct performance comparison between criterion-specific datasets and their corresponding baseline, measuring the impact of each criterion.

\begin{table}[t!]
	\centering
	\resizebox{\columnwidth}{!}{
\begin{tabular}{c|c|c}
\hline 
			Dataset & Included Fields & Number of TRs for Training \\
			\hline
			HTI & H + T + Impact & 8,641 \\
			\hline
			HTF & H + T + Frequency & 11,864 \\
			\hline
			HTC & H + T + Condition & 10,175 \\
			\hline
			HTR & H + T + Reproducibility & 8,097 \\
			\hline
			Single Model & - & 14,504 \\
\hline 
		\end{tabular}
	}
    \caption{Criterion-specific training datasets and sample sizes (number of TRs used for training). Included fields indicate query construction. Abbreviations: H = \textit{headline}, T = \textit{trouble description}, I = \textit{impact}, F = \textit{frequency}, C = \textit{condition}, R = \textit{reproducibility}.}
    \label{datastate_table}
\end{table}

From a subset of the TR corpus that includes all criteria shown in Figure~\ref{tr_example}, we randomly extract two non-overlapping sets of 1,000 TRs. These sets form the basis for the validation and test datasets used in each experiment. It is important to note that the TRs in the validation and test datasets remain constant across all experiments to ensure consistent evaluation. Table~\ref{datastate_table} reports the different datasets along with their metrics.

\subsection{Evaluation Strategy}

To assess the effectiveness of each experiment, we undertake a comprehensive comparison between the performance of each criterion-specific model and its corresponding baseline. This comparative approach enables us to measure the impact of the examined TR observation criteria on the overall performance of the retrieval system, which is calculated for each criterion (impact on system, condition, and others) as follows:
\vspace{-2pt}
\begin{equation}
	I_{C}=P_{C}-P_{C_{baseline}}
\end{equation}
\vspace{-2pt}

Where $P_{C}$ represents the performance of a criterion-specific model for the criterion under study. $P_{C_{baseline}}$denotes the performance of the corresponding baseline model, which includes headline and trouble description, but not the criterion under study. $I_{C}$  is the impact score, indicating the performance difference caused by introducing the specific criterion being evaluated.

The impact score for each criterion is calculated for both bi-encoder and cross-encoder architectures within the two-stage workflow, as shown in Figure~\ref{fig_over}. In addition to this, we compare the overall retrieval performance across criterion-specific models, the CREST ensemble, and the single model baseline. This comparison is conducted under both the full two-stage setup and the cross-encoder (monoTeleRoBERTa) in isolation. The goal of this evaluation is to determine the effectiveness of incorporating criterion-specific signals into the retrieval process and to assess whether the ensemble strategy in CREST leads to consistent performance improvements over individual models and baselines across different configurations.

To assess confidence calibration, we adopt the methodology proposed by \citet{penha2021calibration}, which transforms the ranking task into a multi-class classification problem. For each query, we select the top five documents based on their retrieval scores and apply a softmax function to normalize the scores into probabilities. This reformulation enables the computation of calibration metrics, allowing us to evaluate how well the predicted relevance scores reflect actual relevance likelihoods. We follow the same evaluation setting presented in their study to ensure consistency and comparability in our calibration analysis.

\subsection{Evaluation Metrics}

In our study, we employ three key metrics to quantify the performance of models on datasets introduced in Section~\ref{sec:dataset}: Mean Reciprocal Rank (MRR), Recall@K, and nDCG. 

\emph{Mean Reciprocal Rank (\textit{MRR}):} For each query, the Reciprocal Rank is the inverse of the rank of the first relevant document. For instance, if the first relevant document appears at position 2 in the retrieval system output, the RR is $\frac{1}{2}$. \textit{MRR} is calculated by obtaining the mean over the RR of all queries.

\emph{Recall@K:} Evaluates the model by its effectiveness in retrieving relevant documents within the top-K results without considering their actual rank.

\emph{Normalized Discounted Cumulative Gain (\textit{nDCG}):} Measures the ranking quality of retrieval results, giving a higher score for relevant documents ranked higher in the list.

\emph{Expected Calibration Error (\textit{ECE}):} To evaluate how well the predicted relevance scores reflect the true likelihood of relevance, we employ the Expected Calibration Error. This metric quantifies the difference between predicted confidence and empirical accuracy. The predictions are grouped into $M$ equally spaced bins based on their confidence scores. For each bin $B_m$, the absolute difference is computed between the average predicted confidence $\hat{p}_i$ and the observed accuracy (i.e., the fraction of correct predictions). The final ECE is the weighted average of these differences across all bins, defined as:

\begin{equation}
\mathrm{ECE}=\sum{m=1}^M \frac{\left|B_m\right|}{n}\left|\frac{1}{B_m} \sum_{i \in B_m}\left[\mathbb{I}\left(\hat{y}_i=y_i\right)-\hat{p}_i\right]\right|
\end{equation}

This metric was originally proposed by \citet{naeini2015obtaining} and has been adapted for neural ranking settings in recent works \citep{penha2021calibration, tam2023parameter}. A lower ECE indicates better calibration, meaning that the model's predicted relevance scores are more trustworthy.

\emph{Calibration Diagrams:} These plots visualize the relationship between confidence and accuracy. A perfectly calibrated model aligns with the diagonal line, where confidence matches accuracy. Deviations from this line indicate miscalibration, with underconfidence or overconfidence depending on the direction of the shift.

\section{Results} \label{sec:results}

In this section, we present the evaluation results of the CREST compared to a single criterion-agnostic model TR retrieval model. We also discuss the impact score of each criterion on retrieval performance. The performance is assessed by several key metrics: Recall@K ($R@5$, $R@10$, $R@15$), Mean Reciprocal Rank (\textit{MRR}), and $nDCG@15$. We present the evaluation results for all bi-encoder and cross-encoder models, including their performance within a two-stage workflow and the cross-encoder in isolation.

\subsection{Impact of Each Criterion (RQ1)}

Table~\ref{result_t_1} and Table~\ref{result_t_colbert} present the criterion-specific models alongside their baselines, using TwinRoBERTa and ColRoBERTa for the IR stage, respectively. Table~\ref{result_t_2} presents the criterion-specific models with their baselines using the monoRoBERTa model for the RR stage. The findings from the evaluation demonstrate that criterion-specific models enhance retrieval performance compared to their respective baselines. In the IR stage using TwinRoBERTa model, HTI, HTF, and HTC consistently outperform their corresponding baselines across all metrics. Compared to the baseline single model that uses all available information, these models also show improved performance, with the exception of HTI in $R@5$, where it slightly underperforms. On the other hand, the HTR model performs worse than both its baseline and the single model across all metrics, indicating that this criterion may not contribute as effectively to improving retrieval quality in the IR setting. 

\Soroush{In the IR stage with ColRoBERTa, the criterion-specific models once again outperform their baselines for HTI and HTC. The same holds for HTF, with the exception of the $R@15$ metric, where the baseline performed slightly better. Similar to the TwinRoBERTa results, these models consistently surpass the single baseline model that uses all available information, except in the case of HTF at $R@5$. HTR follows a comparable trend to TwinRoBERTa by performing worse than both its respective baseline and the single baseline model. However, in the ColRoBERTa setting, HTR performs considerably better, and its negative effect is relatively minor. Overall, these outcomes confirm the recurring pattern that HTR can reduce performance when used alone in IR stage, while the other criteria contribute positively in most cases.}

\Soroush{RR results are reported using the top-K candidates produced by the \emph{ColRoBERTa} IR configuration, which demonstrated consistently superior performance and thus provides the most reliable candidate pool for re-ranking.} \Rem{In the RR stage, all criterion-specific models, including HTR, show clear improvements over their respective baselines and the single model. This reinforces the strength of focused representations when used in conjunction with full document-query interactions enabled by the cross-encoder. Notably, the HTR model, which underperformed in the IR stage, achieves competitive results in the RR stage, indicating that the richer context available during re-ranking allows the model to better utilize HTR-specific signals that may have been less effective in the retrieval-only setting.} \Soroush{In the RR stage, all criterion-specific models outperform their baselines and the single model across most metrics. A notable reversal occurs with HTR, which underperforms its baseline in IR under both TwinRoBERTa and ColRoBERTa, yet in RR achieves the highest recall at all cutoffs and ties for the top $nDCG@15$, with only $MRR$ slightly higher for HTF. This suggests that the richer contextual information available during re-ranking enables the model to make better use of HTR-specific signals that were less effective in the retrieval-only setting. HTI and HTC remain strong in RR, but their advantage is smaller than in IR. This pattern suggests that criterion-specific modeling continues to add value, with impact and condition cues driving retrieval performance, while the other criteria help refine the final ranking.}

\begin{table}[!t]
	\centering
    \resizebox{\textwidth}{!}{
	\begin{tabular}{c|cccccccccc} 
		\hline 
		Initial Retrieval  &  \multicolumn{9}{|c}{TwinRoBERTa-base} \\
		(Bi-Encoder) & HTI & HTI baseline & HTF & HTF baseline & HTC & HTC baseline & HTR & HTR baseline & Single Model & BM25 \\
		\hline $R@5$ & $49.85 \%$ & $45.95 \%$ & $52.95 \%$ & $49.25 \%$ & $51.35 \%$ & $51.05 \%$ & $50.15 \%$ & $\mathbf{51.45} \%$ & $49.95 \%$ & $46.85 \%$\\
		\hline $R@10$ & $58.56 \%$ & $53.55 \%$ & $\mathbf{61.36} \%$ & $57.56 \%$ & $59.79 \%$ & $57.96 \%$ & $57.66 \%$ & $59.46 \%$ & $57.96 \%$ & $51.55 \%$ \\
		\hline $R@15$ & $64.36 \%$ & $59.16 \%$ & $\mathbf{65 .77 \%}$ & $62.26 \%$ & $65.17 \%$ & $62.96 \%$ & $61.96 \%$ & $64.96 \%$ & $62.66 \%$ & $54.95 \%$ \\
		\hline $MRR$ & $42.19 \%$ & $38.59 \%$ & $\mathbf{43.89} \%$ & $40.75 \%$ & $43.14 \%$ & $41.71 \%$ & $41.92 \%$ & $42.95 \%$ & $42.04 \%$ & $30.58 \%$ \\
		\hline $nDCG@15$ & $\mathbf{52.87} \%$ & $49.68 \%$ & $48.39 \%$ & $45.08 \%$ & $47.58 \%$ & $46.00 \%$ & $45.95 \%$ & $47.40 \%$ & $46.19 \%$ & $42.73 \%$ \\
		\hline 
	\end{tabular}
	}
    \caption{Performance of criterion-specific models, their baselines, and a single criterion-agnostic model for TwinRoBERTa-base encoder in the IR stage.}
	\label{result_t_1}
\end{table}

\begin{table}[!t]
	\centering
    \resizebox{\textwidth}{!}{
	\begin{tabular}{c|cccccccccc} 
		\hline 
		Initial Retrieval  &  \multicolumn{9}{|c}{ColRoBERTa-base} \\
		(Bi-Encoder) & HTI & HTI baseline & HTF & HTF baseline & HTC & HTC baseline & HTR & HTR baseline & Single Model & BM25 \\
		\hline $R@5$ & $\mathbf{58.16 \%}$ & $56.96 \%$ & $55.06 \%$ & $54.95 \%$ & $56.86 \%$ & $53.85 \%$ & $52.95 \%$ & $53.75 \%$ & $55.56 \%$ & $46.85 \%$ \\
        \hline $R@10$ & $65.76 \%$ & $63.66 \%$ & $63.86 \%$ & $63.36 \%$ & $\mathbf{65.77 \%}$ & $62.26 \%$ & $61.86 \%$ & $61.27 \%$ & $62.66 \%$ & $54.95 \%$ \\
        \hline $R@15$ & $69.37 \%$ & $68.57 \%$ & $67.97 \%$ & $68.27 \%$ & $\mathbf{70.97 \%}$ & $66.97 \%$ & $67.17 \%$ & $66.27 \%$ & $67.17 \%$ & $51.55 \%$ \\
		\hline $MRR$ & $49.07 \%$ & $47.42 \%$ & $47.22 \%$ & $46.93 \%$ & $\mathbf{49.13 \%}$ & $47.09 \%$ & $45.61 \%$ & $46.79 \%$ & $45.98 \%$ & $30.58 \%$ \\
		\hline $nDCG@15$ & $53.25 \%$ & $52 \%$ & $51.45 \%$ & $51.28 \%$ & $\mathbf{53.63 \%}$ & $51.06 \%$ & $50.01 \%$ & $50.70 \%$ & $50.29 \%$ & $42.73 \%$ \\
		\hline 
	\end{tabular}
	}
    \caption{\Soroush{Performance of criterion-specific models, their baselines, and a single criterion-agnostic model for ColRoBERTa-base encoder in the IR stage.}}
	\label{result_t_colbert}
\end{table}

\begin{table}[t]
	\centering
	\resizebox{\textwidth}{!}{
    \begin{tabular}{c|ccccccccc} 
		\hline 
		Re-Ranking   &  \multicolumn{9}{|c}{monoRoBERTa-base} \\
		(Cross-Encoder) & HTI & HTI baseline & HTF & HTF baseline & HTC & HTC baseline & HTR & HTR baseline & Single Model  \\
		\hline $R@5$ & $65.87 \%$ & $60.16 \%$ & $65.47 \%$ & $62.06 \%$ & $66.27 \%$ & $61.56 \%$ & $\mathbf{66.67} \%$ & $60.36 \%$ & $63.66 \%$ \\
		\hline $R@10$ & $74.17 \%$ & $68.77 \%$ & $74.17 \%$ & $70.97 \%$ & $74.67 \%$ & $70.17 \%$ & $\mathbf{75.88} \%$ & $68.77 \%$ & $70.27 \%$ \\
		\hline $R@15$ & $77.78 \%$ & $72.37 \%$ & $78.48 \%$ & $73.67 \%$ & $79.28 \%$ & $73.87 \%$ & $\mathbf{79.58} \%$ & $72.27 \%$ & $74.17 \%$ \\
		\hline $MRR$ & $52.02 \%$ & $49.52 \%$ & $\mathbf{52.88} \%$ & $51.79\%$ & $50.58 \%$ & $51.19 \%$ & $52.42 \%$ & $48.16 \%$ & $51.77 \%$ \\
		\hline $nDCG@15$ & $57.89 \%$ & $54.72 \%$ & $\mathbf{58.66} \%$ & $56.78 \%$ & $57.17 \%$ & $56.40 \%$ & $\mathbf{58.66} \%$ & $53.72 \%$ & $56.93 \%$ \\
		\hline 
	\end{tabular}
    }
    \caption{\Soroush{Performance of criterion-specific models, their baselines, and a single model for the monoTeleBERTa-base encoder in the RR stage.}}
	\label{result_t_2}
\end{table}

Table~\ref{result_t_3} presents the performance of criterion-specific models with their baselines using only the isolated monoRoBERTa cross-encoder, without benefiting from the IR stage. The results reveal that the single model, which utilizes all available criteria, consistently achieves better performance across nearly all metrics and settings. This stands in contrast to the RR stage results, where criterion-specific models surpassed both their baselines and the single model. The lack of improvement here suggests that the re-ranking benefits observed earlier rely heavily on the synergy between the IR and RR stages rather than on the cross-encoder alone. Without the support of an IR stage to pre-select relevant candidates, the cross-encoder operates over a broader, noisier set of inputs, which can dilute the effectiveness of specialized representations. 

Overall, these results emphasize that criterion specialization is most effective when integrated into a pipeline where the IR stage helps isolate more relevant candidates, creating a setting where the RR model can more effectively leverage targeted representations.

\begin{table}[]
	\centering
    \resizebox{\textwidth}{!}{
    	\begin{tabular}{c|ccccccccc} 
    		\hline 
    		Isolated &  \multicolumn{9}{|c}{monoRoBERTa-base} \\
    		(Cross-Encoder) & HTI & HTI baseline & HTF & HTF baseline & HTC & HTC baseline & HTR & HTR baseline & Single Model  \\
    		\hline $R@5$ & $56.06 \%$ & $55.36 \%$ & $55.66 \%$ & $55.86 \%$ & $54.35 \%$ & $54.75 \%$ & $58.46 \%$ & $50.65 \%$ & $\mathbf{60.96} \%$ \\
    		\hline $R@10$ & $67.47 \%$ & $65.27 \%$ & $65.57 \%$ & $65.47 \%$ & $66.47 \%$ & $65.47 \%$ & $67.07 \%$ & $61.56 \%$ & $\mathbf{68.57} \%$ \\
    		\hline $R@15$ & $\mathbf{73.67} \%$ & $70.97 \%$ & $70.57 \%$ & $70.07 \%$ & $72.37 \%$ & $70.57 \%$ & $72.97 \%$ & $67.87 \%$ & $73.27 \%$ \\
    		\hline $MRR$ & $43.76 \%$ & $42.53 \%$ & $43.14 \%$ & $42.69 \%$ & $40.40 \%$ & $42.21 \%$ & $44.68 \%$ & $37.55 \%$ & $\mathbf{47.24} \%$ \\
    		\hline $nDCG@15$ & $50.15 \%$ & $48.60 \%$ & $48.98 \%$ & $48.51 \%$ & $47.29 \%$ & $48.28 \%$ & $50.78 \%$ & $43.94 \%$ & $\mathbf{52.90} \%$ \\
    		\hline 
    	\end{tabular}
	}
    \caption{Performance of criterion-specific models, their baselines, and a single model for the isolated monoTeleBERTa-base encoder.}
	\label{result_t_3}
\end{table}

\Revision{Table~\ref{result_t_4} presents the impact scores of each criterion across both the IR and RR stages. The results indicate a clear hierarchy among the criteria, with ``system impact'' (HTI) showing the strongest positive influence in both stages, followed by ``frequency'' (HTF) and ``condition'' (HTC). In particular, ``reproducibility'' (HTR) shows a negative impact across all IR metrics but demonstrates a positive contribution in the RR stage, particularly in MRR and nDCG@15. , suggesting that this criterion may require deeper context or higher-level interactions to become useful for ranking.}

\Soroush{Table~\ref{result_t_4} presents the impact scores of each criterion across both the IR and RR stages. With TwinRoBERTa in IR, ``system impact'' (HTI) shows the strongest positive influence across all metrics, with ``frequency'' (HTF) and ``condition'' (HTC) following, and ``reproducibility'' (HTR) reducing performance. With ColRoBERTa in IR, ``condition'' emerges as the leading signal while ``system impact'' remains beneficial, ``frequency'' offers smaller gains, and ``reproducibility'' is mixed, hurting early precision but providing slight improvements at deeper recall. In the RR stage using monoRoBERTa over ColRoBERTa candidates, ``reproducibility'' delivers the largest gains on all metrics, while impact and ``frequency'' remain positive and ``condition'' contributes less to MRR.}

\Soroush{One possible reason why HTR performs worse in the IR stage is that the bi-encoder compares short query embeddings with pre-computed document embeddings, which tends to favour concise descriptors. HTR text often includes procedural steps, boilerplate phrases, and local details such as paths or versions, which may dilute the main semantic signal. Because answers do not always mirror these stepwise details, the similarity match is often weaker than for HTI, HTF, or HTC, which contain clearer fault descriptors and affected components. In contrast, the RR stage may benefit more from HTR since the cross-encoder can capture word interactions, ordering, and negation, making it easier to align procedural steps with the answer section resolution explanations.}

\begin{comment} 
\begin{table}[!t]
	\centering
    \normalsize
    \resizebox{\textwidth}{!}{
    	\begin{tabular}{c|cccc|cccc} 
    		\hline 
    		Bi-Encoder  & \multicolumn{4}{|c|}{            IR stage - TwinRoBERTa-base            } & \multicolumn{4}{c}{            RR Stage - monoRoBERTa-base            } \\
    		(TwinBERT) & HTI & HTF & HTC & HTR & HTI & HTF & HTC & HTR \\
    		\hline $R@5$ & $\mathbf{3.9}$ & 3.7 & 0.3 & -1.3 & $\mathbf{7.18}$ & 6.61 & 4.38 & 5.61 \\
    		\hline $R@10$ & $\mathbf{5.01}$ & 3.8 & 1.83 & -1.8 & $\mathbf{6.21}$ & 5.5 & 6.0 & 4.9\\
    		\hline $R@15$ & $\mathbf{5.2}$ & 3.51 & 2.21 & -3 & $\mathbf{7.0}$ & 4.98 & 5.0 & 3.6 \\
    		\hline $MRR$ & $\mathbf{3.6}$ & 3.14 & 1.43 & -1.03 & $4.6$ & 4.19 & 2.46 & $\mathbf{5.62}$ \\
    		\hline $nDCG@15$ & 3.19 & $\mathbf{3.31}$ & 1.58 & -1.45 & $\mathbf{15.22}$ & 4.52 & 3.07 & 5.15 \\
    		\hline 
    	\end{tabular}
    }
    \caption{Impact of each criterion on the retrieval model performance.}
    \label{result_t_4}
\end{table}
\end{comment} 

\begin{table}[!t]
	\centering
    \normalsize
    \resizebox{\textwidth}{!}{
    	\begin{tabular}{c|cccc|cccc|cccc} 
    		\hline 
    		Two Stage  & \multicolumn{4}{|c|}{            IR stage - TwinRoBERTa-base            } & \multicolumn{4}{c|}{            IR stage - ColRoBERTa IR            } & \multicolumn{4}{c}{            RR Stage - monoRoBERTa-base            } \\
    		Retrieval  & HTI & HTF & HTC & HTR & HTI & HTF & HTC & HTR & HTI & HTF & HTC & HTR \\
    		\hline 
            $R@5$       & $\mathbf{3.9}$ & 3.7  & 0.3  & -1.3 & 1.2 & 0.1 & $\mathbf{3.01}$ & -0.8 & 5.71 & 3.41 & 4.71 & $\mathbf{6.31}$ \\
    		\hline 
            $R@10$      & $\mathbf{5.01}$ & 3.8  & 1.83 & -1.8 & 2.1 & 0.50 & $\mathbf{3.51}$ & 0.59 & 5.40 & 3.20 & 4.50 & $\mathbf{7.10}$  \\
    		\hline 
            $R@15$      & $\mathbf{5.2}$ & 3.51 & 2.21 & -3.0 & 0.8 & -0.3 & $\mathbf{4.0}$ & 0.9 & 5.41 & 4.81 & 5.41 & $\mathbf{7.31}$  \\
    		\hline 
            $MRR$       & $\mathbf{3.6}$ & 3.14 & 1.43 & -1.03& 1.65 & 0.29 & $\mathbf{2.04}$ & -1.18 & 2.50 & 1.09 & -0.61 & $\mathbf{4.26}$ \\
    		\hline 
            $nDCG@15$   & 3.19 & $\mathbf{3.31}$ & 1.58 & -1.45& 1.25 & 0.17 & $\mathbf{2.57}$ & -0.69 & 3.17 & 1.88 & 0.77 & $\mathbf{4.94}$ \\
    		\hline 
    	\end{tabular}
    }
    \caption{\Soroush{Impact of each criterion on the retrieval model performance. RR-stage result is computed using monoRoBERTa with \emph{ColRoBERTa} as the preceding IR stage.}}
    \label{result_t_4}
\end{table}

\begin{table}[!t]
\centering
\setlength{\tabcolsep}{4pt}
\footnotesize
\resizebox{\columnwidth}{!}{
\begin{tabular}{l|l|c|c|c|c|c}
\toprule
\textbf{Model} & \textbf{Variant} & \textbf{MRR} & \textbf{R@5} & \textbf{R@10} & \textbf{R@15} & \textbf{nDCG@15} \\
\midrule
\multirow{5}{*}{IR - ColRoBERTa}
  & CREST (All) & $52.50\%$ & $61.36\%$ & $69.27\%$ & $75.18\%$ & $57.19\%$ \\
  & CREST w/o I & $52.20\%$ {\scriptsize($\downarrow 0.30$)} & $60.62\%$ {\scriptsize($\downarrow 0.74$)} & $67.94\%$ {\scriptsize($\downarrow 1.33$)} & $72.14\%$ {\scriptsize($\downarrow 3.04$)} & $56.26\%$ {\scriptsize($\downarrow 0.93$)} \\
  & CREST w/o F & $52.46\%$ {\scriptsize($\downarrow 0.04$)} & $61.66\%$ {\scriptsize($\uparrow 0.30$)} & $68.87\%$ {\scriptsize($\downarrow 0.40$)} & $74.47\%$ {\scriptsize($\downarrow 0.71$)} & $57.00\%$ {\scriptsize($\downarrow 0.19$)} \\
  & CREST w/o C & $51.21\%$ {\scriptsize($\downarrow 1.29$)} & $59.96\%$ {\scriptsize($\downarrow 1.40$)} & $69.27\%$ {\scriptsize($0.00$)} & $73.27\%$ {\scriptsize($\downarrow 1.91$)} & $55.77\%$ {\scriptsize($\downarrow 1.42$)} \\
  & CREST w/o R & $51.87\%$ {\scriptsize($\downarrow 0.63$)} & $60.92\%$ {\scriptsize($\downarrow 0.44$)} & $68.54\%$ {\scriptsize($\downarrow 0.73$)} & $74.05\%$ {\scriptsize($\downarrow 1.13$)} & $56.46\%$ {\scriptsize($\downarrow 0.73$)} \\
\midrule
\multirow{5}{*}{IR - TwinRoBERTa}
  & CREST (All) & $50.64\%$ & $60.02\%$ & $67.74\%$ & $72.04\%$ & $55.07\%$ \\
  & CREST w/o I & $49.53\%$ {\scriptsize($\downarrow 1.11$)} & $59.02\%$ {\scriptsize($\downarrow 1.00$)} & $66.63\%$ {\scriptsize($\downarrow 1.11$)} & $71.14\%$ {\scriptsize($\downarrow 0.90$)} & $53.99\%$ {\scriptsize($\downarrow 1.08$)} \\
  & CREST w/o F & $49.25\%$ {\scriptsize($\downarrow 1.39$)} & $58.02\%$ {\scriptsize($\downarrow 2.00$)} & $64.93\%$ {\scriptsize($\downarrow 2.81$)} & $71.24\%$ {\scriptsize($\downarrow 0.80$)} & $53.75\%$ {\scriptsize($\downarrow 1.32$)} \\
  & CREST w/o C & $49.63\%$ {\scriptsize($\downarrow 1.01$)} & $58.86\%$ {\scriptsize($\downarrow 1.16$)} & $66.47\%$ {\scriptsize($\downarrow 1.27$)} & $71.47\%$ {\scriptsize($\downarrow 0.57$)} & $54.16\%$ {\scriptsize($\downarrow 0.91$)} \\
  & CREST w/o R & $50.14\%$ {\scriptsize($\downarrow 0.50$)} & $58.92\%$ {\scriptsize($\downarrow 1.10$)} & $67.23\%$ {\scriptsize($\downarrow 0.51$)} & $71.44\%$ {\scriptsize($\downarrow 0.60$)} & $54.54\%$ {\scriptsize($\downarrow 0.53$)} \\
\midrule
\multirow{5}{*}{RR - monoRoBERTa}
  & CREST (All) & $57.69\%$ & $70.07\%$ & $77.73\%$ & $81\%$ & $63.08\%$ \\
  & CREST w/o I & $56.62\%$ {\scriptsize($\downarrow 1.06$)} & $69.68\%$ {\scriptsize($\downarrow 0.39$)} & $77.62\%$ {\scriptsize($\downarrow 0.11$)} & $80.87\%$ {\scriptsize($\downarrow 0.13$)} & $62.26\%$ {\scriptsize($\downarrow 0.82$)} \\
  & CREST w/o F & $56.28\%$ {\scriptsize($\downarrow 1.41$)} & $69.99\%$ {\scriptsize($\downarrow 0.08$)} & $77.92\%$ {\scriptsize($\uparrow 0.19$)} & $80.87\%$ {\scriptsize($\downarrow 0.13$)} & $62\%$ {\scriptsize($\downarrow 1.08$)} \\
  & CREST w/o C & $57.24\%$ {\scriptsize($\downarrow 0.45$)} & $69.55\%$ {\scriptsize($\downarrow 0.52$)} & $77.49\%$ {\scriptsize($\downarrow 0.24$)} & $80.75\%$ {\scriptsize($\downarrow 0.25$)} & $62.68\%$ {\scriptsize($\downarrow 0.4$)} \\
  & CREST w/o R & $55.91\%$ {\scriptsize($\downarrow 1.78$)} & $68.43\%$ {\scriptsize($\downarrow 1.64$)} & $76.88\%$ {\scriptsize($\downarrow 0.85$)} & $80.86\%$ {\scriptsize($\downarrow 0.14$)} & $61.66\%$ {\scriptsize($\downarrow 1.42$)} \\
\bottomrule
\end{tabular}}
\vspace{2pt}
\caption{\Soroush{Ablation of criterion-specific models in the CREST ensemble (percent). Differences in parentheses indicate percentage-point change vs.\ the model's \emph{CREST (All)}.  I = Impact (HTI), R = Reproducibility (HTR), F = Frequency (HTF), C = Condition (HTC). “w/o X” excludes criterion X from the ensemble. Differences are percentage points relative to \emph{CREST (All)} for the same backbone.}}
\label{criterion_ablation}
\end{table}

\Soroush{To further analyze the impact of each criterion on ensemble performance, Table~\ref{criterion_ablation} presents results in which CREST is compared with all criteria active and with one removed. The comparison shows that across all three backbones, \emph{CREST (All)} consistently delivers the most reliable performance. Removing any single criterion degrades every evaluation metric, highlighting their strong complementarity. A notable case arises with reproducibility (\emph{R}), which on its own has a negative effect in IR tasks as shown in Table~\ref{result_t_4}. Yet Table~\ref{criterion_ablation} reveals that excluding \emph{R} from the ensemble reduces effectiveness, indicating that CREST is able to exploit its value in combination with other criteria. This suggests that \emph{R} supplies a complementary signal that is not fully captured elsewhere. For both IR models the effect of removing \emph{R} is less severe than removing other criteria, whereas for RR it has a greater negative impact. These patterns suggest that the ensemble leverages \emph{R} as a high-precision discriminator in RR while down-weighting its noisier effect in IR, which explains why its exclusion harms final performance.

Overall, the ablation study demonstrates that each criterion contributes positively when aggregated. By providing distinct forms of evidence, they enhance ranking quality once richer interactions are integrated, supporting the use of the complete criterion set in CREST.} The results also suggest that the contribution of each criterion is not uniform and varies across the IR and RR stages. HTI, HTF, and HTC consistently show a positive impact, highlighting their critical role in both stages of retrieval. Interestingly, HTR, which negatively affects performance in the IR stage, shows notable improvements in the RR stage. This indicates that certain criteria, like HTR, may require a richer interaction or additional context to become useful, which the RR stage is better able to provide. Moreover, it underscores the need to revisit the current TR template structure to ensure that the most influential information is surfacing early, making it easier for readers to quickly identify the most important information.

\subsection{Performance improvement by CREST (RQ2)}

\Rem{CREST was run twice, once with all criteria and then with all criteria except ``reproducibility'', due to its negative impact on performance, as reported in Table~\ref{result_t_4}. The best performance was achieved when all criteria were used, highlighting the model's ability to benefit from every available criterion. Since CREST demonstrated superior performance when all criteria were used, this section will include only results from this scenario.}

Table~\ref{result_t_5} demonstrates the performance improvement of CREST over the single criterion-agnostic retrieval model in the two-stage workflow. \Soroush{This comparison includes results for the IR stage (TwinRoBERTa and ColRoBERTa), followed by the RR stage (monoRoBERTa) over ColRoBERTa candidates.}

\begin{table*}[!t]
	\centering
    \normalsize
    \resizebox{\textwidth}{!}{
    	\begin{tabular}{c|c|c|c|c|c|c}
    		\hline 
    		& \multicolumn{2}{|c|}{ IR (TwinRoBERTa) } &  \multicolumn{2}{|c|}{ IR (ColRoBERTa) } & \multicolumn{2}{|c}{ RR (monoRoBERTa) } \\
    		\hline & CREST & \begin{tabular}{l} 
    			Single criterion-agnostic \\ retrieval model \\
    		\end{tabular} & CREST & \begin{tabular}{l} 
    			Single criterion-agnostic \\ retrieval model \\
    		\end{tabular} & CREST & \begin{tabular}{l} 
    			Single criterion-agnostic \\ retrieval model \\
    		\end{tabular} \\
    		\hline $R@5$ & $60.16 \%$ & $49.95 \%$ & $61.36 \%$ & $55.56 \%$ & $70.07 \%$ & $63.66 \%$ \\
    		\hline $R@10$ & $67.67 \%$ & $57.96 \%$ & $69.27 \%$ & $62.66 \%$ & $77.73 \%$ & $70.27 \%$ \\
    		\hline $R@15$ & $72.67 \%$ & $62.66 \%$ & $75.18 \%$ & $67.17 \%$ & $81 \%$ & $74.17 \%$ \\
    		\hline $MRR$ & $50.21 \%$ & $42.04 \%$ & $52.5 \%$ & $45.98 \%$ & $57.69 \%$ & $51.77 \%$ \\
    		\hline $nDCG@15$ & $54.87 \%$ & $46.19 \%$ & $57.19 \%$ & $50.29 \%$ & $63.08 \%$ & $56.93 \%$ \\
    		\hline 
    	\end{tabular}
	}
    \caption{\Soroush{CREST performance improvement over the single criterion-agnostic model in the two-stage (IR-RR) workflow. RR-stage result is computed using monoRoBERTa with \emph{ColRoBERTa} as the preceding IR stag}}
    \label{result_t_5}
\end{table*}

\Rem{The performance evaluation of CREST demonstrates significant gains across all metrics, compared to a single criterion-agnostic model approach in both the IR and RR stages. In the IR stage, \Rem{using all criterion-specific models (HTI, HTC, HTF, HTR)}by aggregating relevance scores from all criterion-specific models (HTI, HTC, HTF, HTR) through weighted ensembling, CREST achieves over $10.2\%$ improvement for $R@5$ compared to the single model. This indicates a substantial improvement in retrieving relevant documents within the top 5 ranked results. Similarly, CREST achieves $9.7\%$ and $10\% $ improvement for $R@10$ and $R@15$, respectively, compared to the single model. CREST also shows $8.1\%$ improvement in \textit{MRR} and $8.6\%$ improvement in $nDCG@15$. 

In the RR stage, CREST still outperforms the single model in all metrics with a $9.31\%$ improvement for $R@5$, $9.51\%$ for $R@10$, and $10.51\%$ for $R@15$. Compared to the criterion-agnostic single model, \textit{MRR} and $nDCG@15$ show significant increases with CREST, demonstrating $8.31\%$ and $8.75\%$ improvements, respectively. \Rem{Although the improvements in the RR stage are notable, they are not as high as those observed in the IR stage.}hese results show that the benefits of CREST persist across both ranking stages, consistently improving retrieval effectiveness.}

\Soroush{The performance evaluation of CREST demonstrates significant gains across all metrics, compared to a single criterion-agnostic model approach in both the IR and RR stages. In the IR stage, by aggregating relevance scores from all criterion-specific models (HTI, HTC, HTF, HTR) through weighted ensembling, CREST achieves over $10.2\%$ improvement for $R@5$ compared to the single model. This indicates a substantial improvement in retrieving relevant documents within the top 5 ranked results. Similarly, CREST achieves $9.7\%$ and $10.0\%$ improvement for $R@10$ and $R@15$, respectively, compared to the single model. CREST also shows $8.1\%$ improvement in \textit{MRR} and $8.6\%$ improvement in $nDCG@15$. Using ColRoBERTa for IR, CREST remains superior to the single model with gains of $5.8\%$ for $R@5$, $6.6\%$ for $R@10$, $8.0\%$ for $R@15$, $6.5\%$ for \textit{MRR}, and $6.9\%$ for $nDCG@15$.

In the RR stage, CREST continues to outperform the single model across all metrics with a $6.4\%$ improvement for $R@5$, $7.5\%$ for $R@10$, and $6.8\%$ for $R@15$. Compared to the criterion-agnostic single model, \textit{MRR} and $nDCG@15$ also improve with CREST by $5.9\%$ and $6.1\%$, respectively. These results show that the benefits of CREST persist across both ranking stages, consistently improving retrieval effectiveness.}

\Rem{Table~\ref{result_t_6} presents a comparison of CREST performance and a single model with criterion-agnostic focusing only on the cross-encoder (monoRoBERTa) stage, without benefiting from the IR stage output. Despite the performance decline shown in Table~\ref{result_t_3}, for criterion-specific models compared to their baselines and the criterion-agnostic model, aggregating all criterion-specific models (HTI, HTC, HTF, HTR), CREST still outperforms the single criterion-agnostic model. This further demonstrates the effectiveness of CREST in enhancing model performance even under input limit constraints. While the aggregated criterion-specific models perform better than the single criterion-agnostic model, they still do not reach the level of performance seen when using the two-stage workflow. Among all metrics, $R@5$ shows a notable improvement of $2.9\%$, while the improvements for other metrics are minimal. This supports the observation that the loss of critical information, such as ``trouble description'' impacts performance.}

Table~\ref{result_t_6} presents the performance of CREST compared to a single criterion-agnostic model, focusing solely on the cross-encoder (monoRoBERTa) without relying on the IR stage. While Table~\ref{result_t_3} showed limited gains for criterion-specific models in isolation, the ensemble approach in CREST yields clear improvements across all metrics. Notably, $R@5$ improves by $5.47\%$, $R@10$ by $7.08\%$, $R@15$ by $8.29\%$, MRR by $5.44\%$, and $nDCG@15$ by $6.09\%$. This suggests that breaking down the input based on TR criteria and assigning it to expert models is more effective than applying a single model to process the entire TR. Furthermore, it shows that results demonstrate that CREST effectively leverages the strengths of individual criterion-specific models and highlights the benefit of integrating multiple perspectives, even without the aid of the IR stage.

\begin{table}[!t]
	\centering
    \normalsize
	\begin{tabular}{c|c|c}
		\hline 
		\multicolumn{3}{c}{Isolated monoRoBERTa} \\
		\hline
		& CREST & Single criterion-agnostic retrieval model \\
		\hline
		$R@5$ & $\mathbf{66.43\%}$ & 60.96\% \\
		\hline
		$R@10$ & $\mathbf{75.65\% }$& 68.57\% \\
		\hline
		$R@15$ & $\mathbf{81.56\%}$ & 73.27\% \\
		\hline
		$MRR$ & $\mathbf{52.68\%}$ & 47.24\% \\
		\hline
		$nDCG@15$& $\mathbf{58.99\%}$ & 52.90\% \\
		\hline 
	\end{tabular}
    \caption{Performance of CREST and a single retrieval model.}
	\label{result_t_6}
\end{table}

These findings indicate that CREST consistently outperforms the single model approach across all metrics in both the IR and RR stages, underscoring the effectiveness of leveraging criterion-specific models for TR retrieval. By modeling each criterion independently and then aggregating their outputs, CREST can capture diverse and complementary signals that a single model might overlook. This leads to more accurate retrieval of relevant TRs, which can directly benefit the troubleshooting workflow.

A more accurate TR retrieval system helps engineers find previously resolved TRs that are more closely aligned with the current issue, improving the relevance and reliability of the suggested solutions. The improved performance of CREST is directly reflected in the utility of the CREST tool. Alongside retrieving more relevant TRs, the tool provides a breakdown of relevance scores across criteria, offering insight into why each result was selected. This makes the retrieval process more interpretable and actionable. For instance, if a match is primarily driven by ``system impact'' and ``condition'', users can quickly assess its relevance to the new issue. This integration not only supports faster resolution but also helps teams focus on the most critical aspects of a problem, reducing manual effort and improving workflow efficiency.

In practice, this means that CREST doesn't just retrieve better matches, it also supports decision-making by highlighting why those matches were selected. This capability helps engineers respond more efficiently, reduces trial-and-error in solution discovery, and enables more consistent handling of recurring issues.

\Soroush{That said, CREST does not always outperform a single model. Failures typically occur when the IR candidate set excludes the true match, a situation more likely with short or ambiguous queries or with sparse observations. In such cases, a single model may be preferable, as it processes the TR observation holistically, increasing query length and potentially reducing ambiguity by incorporating information overlooked by criterion-specific models. Despite quality assurance measures prior to TR publication, instances still arise where TR creators comply with all requirements yet provide descriptions that lack sufficient detail or contain ambiguity, negatively affecting both CREST and the single model. Another scenario in which a single model is advantageous arises when only a single criterion (for example, ``system impact'') is present for a TR. Criterion-specific models become vulnerable to noise from that single criterion, with no others available to mitigate it. In such situations, a reasonable fallback is to adopt a criterion-agnostic single model.}

\begin{table}[!t]
	\centering
    \normalsize
	\begin{tabular}{c|c|c}
		\hline
		& Initial Retrieval & Re-Ranking \\
		\hline
		HTI & 0.0287 & 0.0186 \\
		\hline
		HTF & 0.0308 & 0.0197 \\
		\hline
		HTC & 0.0304 & 0.0213 \\
		\hline
		HTR & 0.0344 & 0.0199 \\
		\hline
		CREST  & $\mathbf{0.0249}$ & $\mathbf{0.0175}$ \\
        \hline
        Single retrieval model & 0.0345 & 0.0254 \\
		\hline 
	\end{tabular}
    \caption{\Soroush{Expected calibration error (ECE) for all models.}}
	\label{tab:ece-results}
\end{table}

\subsection{Relevance Score Calibration Analysis (RQ3)}

To better understand the quality of the predicted relevance scores and their alignment with actual outcomes, we analyze the calibration of each model using the Expected Calibration Error (ECE). A lower ECE indicates that the predicted probabilities better reflect the true likelihood of relevance, contributing to more trustworthy and reliable relevance scores for each criterion. This is particularly important in retrieval systems where confidence estimates play a role in guiding downstream decisions. Table~\ref{tab:ece-results} reports the ECE values for all models across both the initial retrieval and re-ranking stages.

As shown in Table~\ref{tab:ece-results}, the criterion-specialized models consistently achieve lower ECE scores compared to the single retrieval model, with the difference being more notable in the re-ranking stage. This suggests that modeling relevance per criterion results in more reliable confidence estimation, without the need for additional calibration methods. Even though calibration was not explicitly targeted during training, the specialized models yield better alignment between predicted and actual relevance probabilities.

\begin{figure}[!h]
\centering
\begin{minipage}[b]{0.48\textwidth}
\centering
\includegraphics[width=\textwidth]{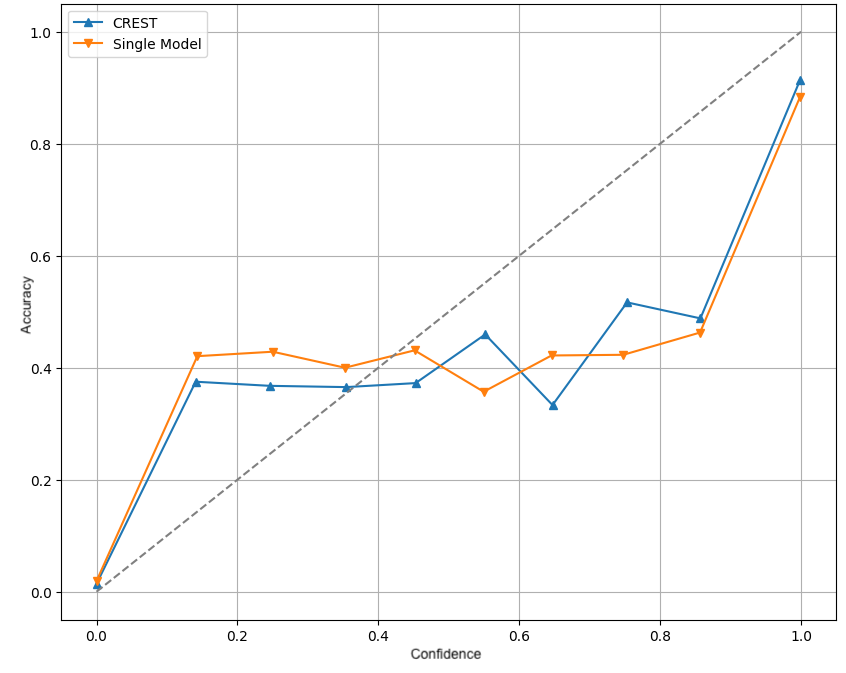}
\end{minipage}
\hfill
\begin{minipage}[b]{0.48\textwidth}
\centering
\includegraphics[width=\textwidth]{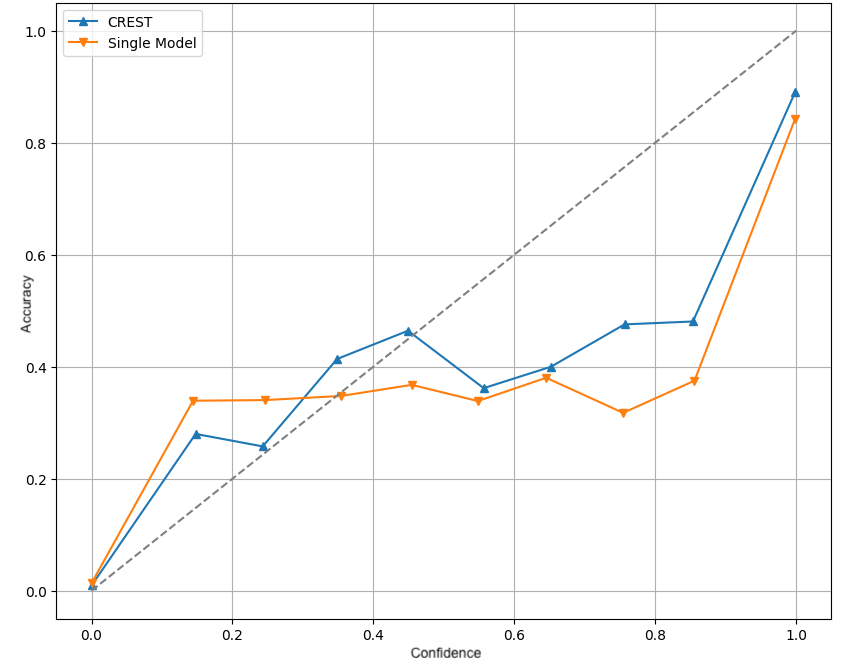}
\end{minipage}
\caption{\Soroush{Calibration diagrams for initial retrieval (left) and re-ranking (right). The left plot shows the calibration performance of the initial retrieval stage, while the right plot illustrates the calibration diagram obtained after re-ranking.}}
\label{fig:calibration_comparison}
\end{figure}

Figure~\ref{fig:calibration_comparison} shows the calibration diagrams for both stages, comparing the CREST model with the single criterion-agnostic retrieval model. In both cases, the ensemble model tracks more closely to the ideal calibration line, while the single model shows larger deviations. These trends are consistent with the ECE results and indicate that combining criterion-specific predictions leads to a better confidence calibration of relevance scores.

\Soroush{This makes the confidence scores easier to interpret, since they can be read as reliable probabilities rather than opaque values. Engineers can then set risk-aware thresholds, decide when to auto-suggest or defer to manual review, and identify which criterion drives a match with dependable certainty. It also lowers the risk of over-confidence and reduces the chance of irrelevant evidence being passed to downstream components such as re-rankers or RAG-style assistants.}

Together, these findings highlight that leveraging criterion-specific models improves not only retrieval performance but also the calibration and reliability of the predicted scores. This is particularly important when transparency in the decision-making process is a key requirement.

\Soroush{
\subsection{Pilot User Study (RQ4)}

We complemented the offline evaluation with a small pilot to check whether the gains we observe translate into practice. The goal was to understand if the criterion-wise scores make the ranking more transparent, whether the recommendations are useful during triage, and how credible and accurate the top results appear to possible end users. 

This study was conducted at Ericsson with three practitioners and five real-life TRs. All participants had QA testing backgrounds: two were actively involved in QA testing at the time of the study, and one was not currently testing but had prior QA experience. One participant had less than five years of experience at Ericsson, and the other two had more than five years.\SoroushN{ For each TR, participants first read its description and then reviewed the top five candidate TRs retrieved by CREST. Each candidate was presented with both the aggregated CREST score and the detailed criterion-wise scores, allowing participants to examine how individual criteria contributed to the final relevance score. They then rated (i) explainability of scoring, (ii) helpfulness for triage tasks (such as root-cause clues, mitigation, symptom matching, and TR authoring), (iii) trustworthiness, and (iv) perceived accuracy, using a six-point scale ranging from “Not at all” to “Excellent”. Participants were given one week to complete the evaluation at their own pace and on average spent about 15 minutes evaluating each TR. At the end of the evaluation, they were also asked to share their overall impressions of CREST. Given the small number of participants and TRs, this pilot should be viewed as exploratory and intended to provide initial qualitative insights.}

\begin{table}[!t]
\centering
\resizebox{\textwidth}{!}{
\begin{tabular}{l|ccc|ccc|ccc|ccc|ccc}
\hline
& \multicolumn{3}{c|}{\textbf{TR1}} & \multicolumn{3}{c|}{\textbf{TR2}} & \multicolumn{3}{c|}{\textbf{TR3}} & \multicolumn{3}{c|}{\textbf{TR4}} & \multicolumn{3}{c}{\textbf{TR5}} \\
\textbf{Participants} & P1 & P2 & P3 & P1 & P2 & P3 & P1 & P2 & P3 & P1 & P2 & P3 & P1 & P2 & P3 \\
\hline
Explainability of Scoring & Good & Good & Good & Good & Fair & Fair & Good & Fair & Very Poor & Good & Fair & Fair & Good & Fair & Good \\
Helpfulness of Ranking     & Good & Good & Very Poor & Good & Fair & Fair & Good & Fair & Good & Good & Good & Good & Excellent & Fair & Good \\
Trustworthiness of Ranking & Good & Fair & Very Poor & Good & Fair & Fair & Good & Good & Fair & Very Poor & Good & Good & Good & Fair & Good \\
Accuracy of Ranking        & Excellent & Fair & Very Poor & Excellent & Good & Good & Excellent & Good & Fair & Very Poor & Good & Good & Good & Fair & Good \\
\hline
\end{tabular}}
\caption{\Soroush{Participants' (P1--P3) ratings for five TRs (TR1--TR5).}}
\label{pivot-user-study}
\end{table}

Table~\ref{pivot-user-study} reports the results of the pilot user study. With 15 ratings per evaluation metric, \emph{explainability of scoring} was ranked as mostly \emph{Good} (8/15, 53\%) or \emph{Fair} (6/15, 40\%), indicating that criterion-wise scores improved transparency, yet leaving room for clearer presentation. \emph{Helpfulness of ranking} was \emph{Good/Excellent} in 10/15 (67\%), suggesting the provided TR lists often surfaced actionable cues. \emph{Trustworthiness} was perceived as \emph{Good} in 8/15 (53\%) and \emph{Fair} in 5/15 (33\%). Lastly, perceived \emph{accuracy} was \emph{Good/Excellent} in 10/15 (67\%), with a minority of \emph{Very Poor} judgments (2/15). We did not observe any ``Not at all'' ratings for any metric.

Differences across TRs were mainly explained by input quality, with sparse or unclear criteria affecting CREST's performance. Findings consistent with both performance and calibration gains observed in offline experiments. Moreover, explainability benefited from the criterion-wise scores, which participants used to interpret why a given TR surfaced and to justify keeping or discarding specific candidates.

Participants feedback on negative cases reinforces our earlier findings, highlighting the strength of CREST when the criteria are informative and its weaker performance when they are brief or ambiguous. Moreover, participant P3 noted that the scoring explanations allowed them to justify selecting the most relevant TR from the candidate list provided, illustrating how transparency in the rankings can support more informed decision-making.  
}

\section{Discussion}

We now discuss the implications of our work (Section~\ref{sec:implications}) and address the threats to validity and how we mitigated them (Section~\ref{sec:ttv}). 

\subsection{Implications} \label{sec:implications}

The proposed CREST model has practical implications for improving the efficiency and reliability of quality assurance (QA) and design teams at Ericsson. By offering a more accurate and interpretable retrieval of relevant trouble reports (TRs), CREST can assist engineers in identifying related past issues more effectively, thereby accelerating the fault resolution process. This is particularly valuable in complex systems where understanding the context and history of software faults plays a crucial role in diagnosing and resolving new incidents.

One of the key benefits of CREST is its ability to surface criterion-specific relevance scores, helping users understand which aspects of a TR (e.g., functional area, fault type, or impacted component) contributed most to the retrieval outcome. This transparency can guide engineers in validating the retrieved TRs, increase their trust in the system, and potentially uncover new resolution strategies based on previously overlooked criteria.

From a practical standpoint, CREST can be integrated into existing TR retrieval workflows within Ericsson with minimal disruption. Depending on the requirements of the task, CREST can be used in the Initial Retrieval stage, especially for latency-sensitive applications, or extended in both stages of a two-stage pipeline to fully leverage its benefits. In addition to its role in TR retrieval, CREST can also serve as a retrieval engine for Retrieval-Augmented Generation (RAG) systems tailored to TR-related tasks. In this context, CREST's interpretable retrieval helps explain why specific information sources were selected and how they contributed to the final output, enhancing both the transparency and reliability of the generated responses.

Overall, CREST offers a practical and effective enhancement to existing TR retrieval systems, enabling Ericsson teams to reduce resolution times, improve traceability, and support more informed decision-making during software maintenance and troubleshooting.

\subsection{Threats to Validity} \label{sec:ttv}

Several factors may threaten the validity of our findings. First, the internal data from Ericsson is proprietary and cannot be publicly shared due to non-disclosure obligations. This data may not generalize to other telecommunication environments, limiting external validity due to different TR characteristics and structures. \Rem{We can not share the replication package as it is proprietary to Ericsson, and their policy restricts the distribution of the code outside the organization.}

Second, the quality of TRs and parsing tools is critical and can impact model performance, especially during deployment. Inconsistencies in these tools or data can influence the effectiveness of the TR retrieval system.

Third, criterion-specific datasets may not contain TRs of similar quality, resulting in performance variations. We introduced a baseline to isolate the effect of each criterion, but differences may still affect results.

\Rem{Fourth, while CREST is designed to enhance interpretability by providing per-criterion relevance scores, we did not explicitly measure how this impacts users' understanding or trust in the retrieval results. Therefore, the extent to which CREST contributes to practical interpretability remains an open question. Future work should include user studies or quantitative analyses to assess this aspect. Additionally, the definition of interpretability is broad. In this work, we use it to refer to the model's ability to provide insight into why specific TRs are recommended, specifically by showing how each TR criterion contributes to the retrieval result.}\Soroush{Fourth, CREST is designed to enhance explainability by exposing per-criterion relevance scores. To assess its practical value, a pilot user study was conducted. The results suggest that per-criterion scores support sense-making and traceability of rankings, highlighting CREST's potential for interpretability and transparency. However, the study was limited in scale and scope, and a larger, more comprehensive evaluation involving broader user groups is needed and left as a possible future direction.}

Moreover, due to Ericsson's policy and resource limitation challenge, we have only evaluated our approach with an internally trained RoBERTa model, and as a result, the impact of the CREST approach may vary across different large language models (LLMs). Additionally, the effectiveness of the CREST approach may not be consistent across various TR retrieval applications, which limits the generalizability of our findings.

While we acknowledge these limitations, we hope to address them in future research to validate and generalize our findings.

\section{Conclusion} \label{sec:conclusion}

This study investigates the impact of various trouble report criteria on the performance of the Initial Retrieval (IR) and Re-Ranking (RR) stages within the TR retrieval system. By utilizing a bi-encoder in the IR stage and a cross-encoder in the RR stage, we were able to evaluate each criterion's influence in a comprehensive two-stage workflow. Notably, criteria such as ``system impact" (HTI) significantly improved recall and ranking metrics during the IR stage, whereas ``reproducibility" (HTR) \Rem{had a detrimental effect on these same metrics} negatively influenced the IR stage but showed positive effects in the RR stage, highlighting that different stages benefit from different types of information. This illustrates the importance of selectively parsing and utilizing specific information to enhance retrieval performance, as opposed to leveraging all available data indiscriminately. \Rem{Additionally, we evaluated the impact of each criterion using a single cross-encoder in isolation. The results revealed that performance declined likely due to the input length limitations inherent to cross-encoders, which are more severe compared to bi-encoders.}The standalone evaluation of each criterion's impact also highlighted the benefits of criterion-specific modeling, especially in re-ranking, where the detailed reasoning capabilities of cross-encoders are better suited to exploit criterion-specific signals.

The proposed Criteria-specific Retrieval via Ensemble of Specialized TR models (CREST) demonstrates a significant advancement in TR retrieval approaches. By training each model within the ensemble to focus on a unique criterion, we cultivated specialization in handling specific types of information. This specialization allowed CREST to effectively combine the diverse strengths of each model, leading to improved overall performance. The findings of this study demonstrate that CREST consistently outperforms single-model approaches across key metrics in both IR and RR stages when applied in a two-stage workflow. \Rem{Furthermore, CREST also showed superior performance in isolated cross-encoder settings compared to single-model approaches, although the performance did not reach the levels achieved with the two-stage workflow.} In addition, even when evaluated in isolation (i.e., without the IR stage), CREST continued to outperform the single-model approach across all metrics, reinforcing the strength of criterion-specific modeling in constrained input settings. Moreover, CREST improves the calibration of predicted confidence scores, resulting in outputs that are not only more precise but also more reliable, which is an important factor when transparency and interpretability are critical in decision-making processes. \Soroush{Finally, the pilot user study demonstrates that CREST's criterion-wise scores can improve perceived transparency and that the recommendations were often judged useful, credible, and accurate for triage, with some responses also suggesting areas for further refinement.}

In summary, CREST substantially enhances the accuracy and efficiency of TR retrieval systems on our industrial dataset. Beyond improved retrieval accuracy, the CREST tool can also support practical decision-making by providing criterion-wise relevance scores for each retrieved TR, making the retrieval process interpretable and traceable. This capability enables engineers to better understand the match rationale, prioritize investigation based on critical factors, and ultimately accelerate issue resolution. Future research should aim to refine criteria aggregation methods and explore the integration of non-TR sources and non-textual information to further elevate TR retrieval system performance. These advancements hold the potential to significantly improve the capabilities and effectiveness of TR retrieval systems in industrial applications.

% \begin{thebibliography}{00}
\bibliographystyle{elsarticle-harv} 
\bibliography{sample-base}

\begin{thebibliography}{50}
\expandafter\ifx\csname natexlab\endcsname\relax\def\natexlab#1{#1}\fi
\providecommand{\url}[1]{\texttt{#1}}
\providecommand{\href}[2]{#2}
\providecommand{\path}[1]{#1}
\providecommand{\DOIprefix}{doi:}
\providecommand{\ArXivprefix}{arXiv:}
\providecommand{\URLprefix}{URL: }
\providecommand{\Pubmedprefix}{pmid:}
\providecommand{\doi}[1]{\href{http://dx.doi.org/#1}{\path{#1}}}
\providecommand{\Pubmed}[1]{\href{pmid:#1}{\path{#1}}}
\providecommand{\bibinfo}[2]{#2}
\ifx\xfnm\relax \def\xfnm[#1]{\unskip,\space#1}\fi
%Type = Article
\bibitem[{Achiam et~al.(2023)Achiam, Adler, Agarwal, Ahmad, Akkaya, Aleman,
  Almeida, Altenschmidt, Altman, Anadkat et~al.}]{achiam2023gpt}
\bibinfo{author}{Achiam, J.}, \bibinfo{author}{Adler, S.},
  \bibinfo{author}{Agarwal, S.}, \bibinfo{author}{Ahmad, L.},
  \bibinfo{author}{Akkaya, I.}, \bibinfo{author}{Aleman, F.L.},
  \bibinfo{author}{Almeida, D.}, \bibinfo{author}{Altenschmidt, J.},
  \bibinfo{author}{Altman, S.}, \bibinfo{author}{Anadkat, S.}, et~al.,
  \bibinfo{year}{2023}.
\newblock \bibinfo{title}{Gpt-4 technical report}.
\newblock \bibinfo{journal}{arXiv preprint arXiv:2303.08774} .
%Type = Article
\bibitem[{Aggarwal et~al.(2021)Aggarwal, Mittal and Battineni}]{b22}
\bibinfo{author}{Aggarwal, A.}, \bibinfo{author}{Mittal, M.},
  \bibinfo{author}{Battineni, G.}, \bibinfo{year}{2021}.
\newblock \bibinfo{title}{Generative adversarial network: An overview of theory
  and applications}.
\newblock \bibinfo{journal}{International Journal of Information Management
  Data Insights} \bibinfo{volume}{1}, \bibinfo{pages}{100004}.
\newblock \DOIprefix\doi{10.1016/j.jjimei.2020.100004}.
%Type = Article
\bibitem[{Anand et~al.(2022)Anand, Lyu, Idahl, Wang, Wallat and
  Zhang}]{anand2022explainable}
\bibinfo{author}{Anand, A.}, \bibinfo{author}{Lyu, L.}, \bibinfo{author}{Idahl,
  M.}, \bibinfo{author}{Wang, Y.}, \bibinfo{author}{Wallat, J.},
  \bibinfo{author}{Zhang, Z.}, \bibinfo{year}{2022}.
\newblock \bibinfo{title}{Explainable information retrieval: A survey}.
\newblock \bibinfo{journal}{arXiv preprint arXiv:2211.02405} .
%Type = Inproceedings
\bibitem[{Bosch et~al.(2022)Bosch, Shalmashi, Yaghoubi, Holm, Gaim and
  Payberah}]{b5}
\bibinfo{author}{Bosch, N.}, \bibinfo{author}{Shalmashi, S.},
  \bibinfo{author}{Yaghoubi, F.}, \bibinfo{author}{Holm, H.},
  \bibinfo{author}{Gaim, F.}, \bibinfo{author}{Payberah, A.H.},
  \bibinfo{year}{2022}.
\newblock \bibinfo{title}{Fine-tuning bert-based language models for duplicate
  trouble report retrieval}, in: \bibinfo{booktitle}{2022 IEEE International
  Conference on Big Data (Big Data)}, pp. \bibinfo{pages}{4737--4745}.
%Type = Book
\bibitem[{Chicco(2021)}]{b11}
\bibinfo{author}{Chicco, D.}, \bibinfo{year}{2021}.
\newblock \bibinfo{title}{Siamese Neural Networks: An Overview}.
\newblock \bibinfo{publisher}{US}, \bibinfo{address}{New York, NY}.
\newblock \DOIprefix\doi{10.1007/978-1-0716-0826-5\_3}.
%Type = Inproceedings
\bibitem[{Devlin et~al.(2019)Devlin, Chang, Lee and Toutanova}]{devlin2019bert}
\bibinfo{author}{Devlin, J.}, \bibinfo{author}{Chang, M.W.},
  \bibinfo{author}{Lee, K.}, \bibinfo{author}{Toutanova, K.},
  \bibinfo{year}{2019}.
\newblock \bibinfo{title}{Bert: Pre-training of deep bidirectional transformers
  for language understanding}, in: \bibinfo{booktitle}{Proceedings of the 2019
  conference of the North American chapter of the association for computational
  linguistics: human language technologies, volume 1 (long and short papers)},
  pp. \bibinfo{pages}{4171--4186}.
%Type = Article
\bibitem[{Grattafiori et~al.(2024)Grattafiori, Dubey, Jauhri, Pandey, Kadian,
  Al-Dahle, Letman, Mathur, Schelten, Vaughan et~al.}]{grattafiori2024llama}
\bibinfo{author}{Grattafiori, A.}, \bibinfo{author}{Dubey, A.},
  \bibinfo{author}{Jauhri, A.}, \bibinfo{author}{Pandey, A.},
  \bibinfo{author}{Kadian, A.}, \bibinfo{author}{Al-Dahle, A.},
  \bibinfo{author}{Letman, A.}, \bibinfo{author}{Mathur, A.},
  \bibinfo{author}{Schelten, A.}, \bibinfo{author}{Vaughan, A.}, et~al.,
  \bibinfo{year}{2024}.
\newblock \bibinfo{title}{The llama 3 herd of models}.
\newblock \bibinfo{journal}{arXiv preprint arXiv:2407.21783} .
%Type = Article
\bibitem[{Grimalt et~al.(2022)Grimalt, Shalmashi, Yaghoubi, Jonsson and
  Payberah}]{b2}
\bibinfo{author}{Grimalt, N.M.I.}, \bibinfo{author}{Shalmashi, S.},
  \bibinfo{author}{Yaghoubi, F.}, \bibinfo{author}{Jonsson, L.},
  \bibinfo{author}{Payberah, A.H.}, \bibinfo{year}{2022}.
\newblock \bibinfo{title}{Berticsson: A recommender system for
  troubleshooting}.
\newblock \bibinfo{journal}{SDU@ AAAI} .
%Type = Inproceedings
\bibitem[{Gururangan et~al.(2022)Gururangan, Lewis, Holtzman, Smith and
  Zettlemoyer}]{b30}
\bibinfo{author}{Gururangan, S.}, \bibinfo{author}{Lewis, M.},
  \bibinfo{author}{Holtzman, A.}, \bibinfo{author}{Smith, N.A.},
  \bibinfo{author}{Zettlemoyer, L.}, \bibinfo{year}{2022}.
\newblock \bibinfo{title}{Demix layers: Disentangling domains for modular
  language modeling}, in: \bibinfo{booktitle}{Proceedings of the 2022
  Conference of the North American Chapter of the Association for Computational
  Linguistics: Human Language Technologies}, pp. \bibinfo{pages}{5557--5576}.
%Type = Masterthesis
\bibitem[{Holm(2021)}]{b1}
\bibinfo{author}{Holm, H.}, \bibinfo{year}{2021}.
\newblock \bibinfo{title}{Bidirectional encoder representations from
  transformers (bert) for question answering in the telecom domain}.
\newblock Master's thesis. KTH, School of Electrical Engineering and Computer
  Science (EECS.
%Type = Inproceedings
\bibitem[{Humeau et~al.(2020)Humeau, Shuster, Lachaux and Weston}]{b23}
\bibinfo{author}{Humeau, S.}, \bibinfo{author}{Shuster, K.},
  \bibinfo{author}{Lachaux, M.A.}, \bibinfo{author}{Weston, J.},
  \bibinfo{year}{2020}.
\newblock \bibinfo{title}{Poly-encoders: Architectures and pre-training
  strategies for fast and accurate multi-sentence scoring}, in:
  \bibinfo{booktitle}{Proceeding of 8th International Conference on Learning
  Representations, 2020}.
%Type = Misc
\bibitem[{Jiang et~al.(2023)Jiang, Sablayrolles, Mensch, Bamford, Chaplot,
  de~las Casas, Bressand, Lengyel, Lample, Saulnier, Lavaud, Lachaux, Stock,
  Scao, Lavril, Wang, Lacroix and Sayed}]{jiang2023mistral7b}
\bibinfo{author}{Jiang, A.Q.}, \bibinfo{author}{Sablayrolles, A.},
  \bibinfo{author}{Mensch, A.}, \bibinfo{author}{Bamford, C.},
  \bibinfo{author}{Chaplot, D.S.}, \bibinfo{author}{de~las Casas, D.},
  \bibinfo{author}{Bressand, F.}, \bibinfo{author}{Lengyel, G.},
  \bibinfo{author}{Lample, G.}, \bibinfo{author}{Saulnier, L.},
  \bibinfo{author}{Lavaud, L.R.}, \bibinfo{author}{Lachaux, M.A.},
  \bibinfo{author}{Stock, P.}, \bibinfo{author}{Scao, T.L.},
  \bibinfo{author}{Lavril, T.}, \bibinfo{author}{Wang, T.},
  \bibinfo{author}{Lacroix, T.}, \bibinfo{author}{Sayed, W.E.},
  \bibinfo{year}{2023}.
\newblock \bibinfo{title}{Mistral 7b}.
\newblock \URLprefix \url{https://arxiv.org/abs/2310.06825},
  \href{http://arxiv.org/abs/2310.06825}{{\tt arXiv:2310.06825}}.
%Type = Inproceedings
\bibitem[{Jung et~al.(2022)Jung, Choi and Rhee}]{b19}
\bibinfo{author}{Jung, E.}, \bibinfo{author}{Choi, J.}, \bibinfo{author}{Rhee,
  W.}, \bibinfo{year}{2022}.
\newblock \bibinfo{title}{Semi-siamese bi-encoder neural ranking model using
  lightweight fine-tuning}, in: \bibinfo{booktitle}{Proceedings of the ACM Web
  Conference 2022}, pp. \bibinfo{pages}{502--511}.
\newblock \DOIprefix\doi{10.1145/3485447.3511978}.
%Type = Inproceedings
\bibitem[{Karapantelakis et~al.(2024)Karapantelakis, Thakur, Nikou, Moradi,
  Olrog, Gaim, Holm, Nimara and Huang}]{karapantelakis2024using}
\bibinfo{author}{Karapantelakis, A.}, \bibinfo{author}{Thakur, M.},
  \bibinfo{author}{Nikou, A.}, \bibinfo{author}{Moradi, F.},
  \bibinfo{author}{Olrog, C.}, \bibinfo{author}{Gaim, F.},
  \bibinfo{author}{Holm, H.}, \bibinfo{author}{Nimara, D.D.},
  \bibinfo{author}{Huang, V.}, \bibinfo{year}{2024}.
\newblock \bibinfo{title}{Using large language models to understand telecom
  standards}, in: \bibinfo{booktitle}{2024 IEEE International Conference on
  Machine Learning for Communication and Networking (ICMLCN)},
  \bibinfo{organization}{IEEE}. pp. \bibinfo{pages}{440--446}.
%Type = Inproceedings
\bibitem[{Khattab and Zaharia(2020)}]{b12}
\bibinfo{author}{Khattab, O.}, \bibinfo{author}{Zaharia, M.},
  \bibinfo{year}{2020}.
\newblock \bibinfo{title}{Colbert: Efficient and effective passage search via
  contextualized late interaction over bert}, in:
  \bibinfo{booktitle}{Proceedings of the 43rd International ACM SIGIR
  Conference on Research and Development in Information Retrieval, 2020}, pp.
  \bibinfo{pages}{39--48}.
\newblock \DOIprefix\doi{10.1145/3397271.3401075}.
%Type = Article
\bibitem[{Kirkpatrick et~al.(2017)Kirkpatrick, Pascanu, Rabinowitz, Veness,
  Desjardins, Rusu, Milan, Quan, Ramalho, Grabska-Barwinska et~al.}]{b15}
\bibinfo{author}{Kirkpatrick, J.}, \bibinfo{author}{Pascanu, R.},
  \bibinfo{author}{Rabinowitz, N.}, \bibinfo{author}{Veness, J.},
  \bibinfo{author}{Desjardins, G.}, \bibinfo{author}{Rusu, A.A.},
  \bibinfo{author}{Milan, K.}, \bibinfo{author}{Quan, J.},
  \bibinfo{author}{Ramalho, T.}, \bibinfo{author}{Grabska-Barwinska, A.},
  et~al., \bibinfo{year}{2017}.
\newblock \bibinfo{title}{Overcoming catastrophic forgetting in neural
  networks}.
\newblock \bibinfo{journal}{Proceedings of the national academy of sciences}
  \bibinfo{volume}{114}, \bibinfo{pages}{3521--3526}.
%Type = Article
\bibitem[{Leonhardt et~al.(2023)Leonhardt, Rudra and
  Anand}]{leonhardt2023extractive}
\bibinfo{author}{Leonhardt, J.}, \bibinfo{author}{Rudra, K.},
  \bibinfo{author}{Anand, A.}, \bibinfo{year}{2023}.
\newblock \bibinfo{title}{Extractive explanations for interpretable text
  ranking}.
\newblock \bibinfo{journal}{ACM Transactions on Information Systems}
  \bibinfo{volume}{41}, \bibinfo{pages}{1--31}.
%Type = Inproceedings
\bibitem[{Li et~al.(2022)Li, Gururangan, Dettmers, Lewis, Althoff, Smith and
  Zettlemoyer}]{b29}
\bibinfo{author}{Li, M.}, \bibinfo{author}{Gururangan, S.},
  \bibinfo{author}{Dettmers, T.}, \bibinfo{author}{Lewis, M.},
  \bibinfo{author}{Althoff, T.}, \bibinfo{author}{Smith, N.A.},
  \bibinfo{author}{Zettlemoyer, L.}, \bibinfo{year}{2022}.
\newblock \bibinfo{title}{Branch-train-merge: Embarrassingly parallel training
  of expert language models}, in: \bibinfo{booktitle}{First Workshop on
  Interpolation Regularizers and Beyond at NeurIPS 2022}.
%Type = Article
\bibitem[{Liu et~al.(2024)Liu, Feng, Xue, Wang, Wu, Lu, Zhao, Deng, Zhang, Ruan
  et~al.}]{liu2024deepseek}
\bibinfo{author}{Liu, A.}, \bibinfo{author}{Feng, B.}, \bibinfo{author}{Xue,
  B.}, \bibinfo{author}{Wang, B.}, \bibinfo{author}{Wu, B.},
  \bibinfo{author}{Lu, C.}, \bibinfo{author}{Zhao, C.}, \bibinfo{author}{Deng,
  C.}, \bibinfo{author}{Zhang, C.}, \bibinfo{author}{Ruan, C.}, et~al.,
  \bibinfo{year}{2024}.
\newblock \bibinfo{title}{Deepseek-v3 technical report}.
\newblock \bibinfo{journal}{arXiv preprint arXiv:2412.19437} .
%Type = Unpublished
\bibitem[{Liu et~al.(2019)Liu, Ott, Goyal, Du, Joshi, Chen, Levy, Lewis,
  Zettlemoyer and Stoyanov}]{b6}
\bibinfo{author}{Liu, Y.}, \bibinfo{author}{Ott, M.}, \bibinfo{author}{Goyal,
  N.}, \bibinfo{author}{Du, J.}, \bibinfo{author}{Joshi, M.},
  \bibinfo{author}{Chen, D.}, \bibinfo{author}{Levy, O.},
  \bibinfo{author}{Lewis, M.}, \bibinfo{author}{Zettlemoyer, L.},
  \bibinfo{author}{Stoyanov, V.}, \bibinfo{year}{2019}.
\newblock \bibinfo{title}{Roberta: A robustly optimized bert pretraining
  approach}.
\newblock \href{http://arxiv.org/abs/1907.11692}{{\tt arXiv:1907.11692}}.
  \bibinfo{note}{arXiv preprint}.
%Type = Inproceedings
\bibitem[{Lu et~al.(2020)Lu, Jiao and Zhang}]{b10}
\bibinfo{author}{Lu, W.}, \bibinfo{author}{Jiao, J.}, \bibinfo{author}{Zhang,
  R.}, \bibinfo{year}{2020}.
\newblock \bibinfo{title}{Twinbert: Distilling knowledge to twin-structured
  compressed bert models for large-scale retrieval}, in:
  \bibinfo{booktitle}{Proceedings of the 29th ACM International Conference on
  Information \& Knowledge Management, 2020}, pp. \bibinfo{pages}{2645--2652}.
%Type = Inproceedings
\bibitem[{Lucchese et~al.(2023)Lucchese, Minello, Nardini, Orlando, Perego and
  Veneri}]{lucchese2023can}
\bibinfo{author}{Lucchese, C.}, \bibinfo{author}{Minello, G.},
  \bibinfo{author}{Nardini, F.M.}, \bibinfo{author}{Orlando, S.},
  \bibinfo{author}{Perego, R.}, \bibinfo{author}{Veneri, A.},
  \bibinfo{year}{2023}.
\newblock \bibinfo{title}{Can embeddings analysis explain large language model
  ranking?}, in: \bibinfo{booktitle}{Proceedings of the 32nd ACM International
  Conference on Information and Knowledge Management}, pp.
  \bibinfo{pages}{4150--4154}.
%Type = Inproceedings
\bibitem[{Ma et~al.(2024)Ma, Wang, Yang, Wei and Lin}]{b27}
\bibinfo{author}{Ma, X.}, \bibinfo{author}{Wang, L.}, \bibinfo{author}{Yang,
  N.}, \bibinfo{author}{Wei, F.}, \bibinfo{author}{Lin, J.},
  \bibinfo{year}{2024}.
\newblock \bibinfo{title}{Fine-tuning llama for multi-stage text retrieval},
  in: \bibinfo{booktitle}{Proceedings Of The 47th International ACM SIGIR
  Conference On Research And Development In Information Retrieval}, pp.
  \bibinfo{pages}{2421--2425}.
\newblock \DOIprefix\doi{10.1145/3626772.3657951}.
%Type = Inproceedings
\bibitem[{Naeini et~al.(2015)Naeini, Cooper and
  Hauskrecht}]{naeini2015obtaining}
\bibinfo{author}{Naeini, M.P.}, \bibinfo{author}{Cooper, G.},
  \bibinfo{author}{Hauskrecht, M.}, \bibinfo{year}{2015}.
\newblock \bibinfo{title}{Obtaining well calibrated probabilities using
  bayesian binning}, in: \bibinfo{booktitle}{Proceedings of the AAAI conference
  on artificial intelligence}.
%Type = Inproceedings
\bibitem[{Nimara et~al.(2024)Nimara, Gebre and Huang}]{nimara2024entity}
\bibinfo{author}{Nimara, D.D.}, \bibinfo{author}{Gebre, F.G.},
  \bibinfo{author}{Huang, V.}, \bibinfo{year}{2024}.
\newblock \bibinfo{title}{Entity recognition in telecommunications using
  domain-adapted language models}, in: \bibinfo{booktitle}{2024 IEEE
  International Conference on Machine Learning for Communication and Networking
  (ICMLCN)}, \bibinfo{organization}{IEEE}. pp. \bibinfo{pages}{240--245}.
%Type = Unpublished
\bibitem[{Nogueira and Cho(2019)}]{b4}
\bibinfo{author}{Nogueira, R.}, \bibinfo{author}{Cho, K.},
  \bibinfo{year}{2019}.
\newblock \bibinfo{title}{Passage re-ranking with bert}.
\newblock \href{http://arxiv.org/abs/1901.04085}{{\tt arXiv:1901.04085}}.
  \bibinfo{note}{arXiv preprint}.
%Type = Unpublished
\bibitem[{Nogueira et~al.(2019)Nogueira, Yang, Cho and Lin}]{b9}
\bibinfo{author}{Nogueira, R.}, \bibinfo{author}{Yang, W.},
  \bibinfo{author}{Cho, K.}, \bibinfo{author}{Lin, J.}, \bibinfo{year}{2019}.
\newblock \bibinfo{title}{Multi-stage document ranking with bert}.
\newblock \href{http://arxiv.org/abs/1910.14424}{{\tt arXiv:1910.14424}}.
  \bibinfo{note}{arXiv preprint}.
%Type = Inproceedings
\bibitem[{Penha and Hauff(2021a)}]{penha-hauff-2021-calibration}
\bibinfo{author}{Penha, G.}, \bibinfo{author}{Hauff, C.},
  \bibinfo{year}{2021}a.
\newblock \bibinfo{title}{On the calibration and uncertainty of neural learning
  to rank models for conversational search}, in: \bibinfo{editor}{Merlo, P.},
  \bibinfo{editor}{Tiedemann, J.}, \bibinfo{editor}{Tsarfaty, R.} (Eds.),
  \bibinfo{booktitle}{Proceedings of the 16th Conference of the European
  Chapter of the Association for Computational Linguistics: Main Volume},
  \bibinfo{publisher}{Association for Computational Linguistics},
  \bibinfo{address}{Online}. pp. \bibinfo{pages}{160--170}.
\newblock \URLprefix \url{https://aclanthology.org/2021.eacl-main.12/},
  \DOIprefix\doi{10.18653/v1/2021.eacl-main.12}.
%Type = Inproceedings
\bibitem[{Penha and Hauff(2021b)}]{penha2021calibration}
\bibinfo{author}{Penha, G.}, \bibinfo{author}{Hauff, C.},
  \bibinfo{year}{2021}b.
\newblock \bibinfo{title}{On the calibration and uncertainty of neural learning
  to rank models for conversational search}, in:
  \bibinfo{booktitle}{Proceedings of the 16th Conference of the European
  Chapter of the Association for Computational Linguistics: Main Volume}, pp.
  \bibinfo{pages}{160--170}.
%Type = Inproceedings
\bibitem[{Puthenputhussery et~al.(2025)Puthenputhussery, Kang, Magnani, Zhang,
  Shang, Yadav, Chandran, Madhani, Fu, Wang et~al.}]{puthenputhussery2025large}
\bibinfo{author}{Puthenputhussery, A.}, \bibinfo{author}{Kang, C.},
  \bibinfo{author}{Magnani, A.}, \bibinfo{author}{Zhang, T.},
  \bibinfo{author}{Shang, H.}, \bibinfo{author}{Yadav, N.},
  \bibinfo{author}{Chandran, P.}, \bibinfo{author}{Madhani, B.},
  \bibinfo{author}{Fu, Y.T.}, \bibinfo{author}{Wang, H.}, et~al.,
  \bibinfo{year}{2025}.
\newblock \bibinfo{title}{Large scale deployment of bert based cross encoder
  model for re-ranking in walmart search engine}, in:
  \bibinfo{booktitle}{Proceedings of the 48th International ACM SIGIR
  Conference on Research and Development in Information Retrieval}, pp.
  \bibinfo{pages}{4365--4369}.
%Type = Article
\bibitem[{Raffel et~al.(2020)Raffel, Shazeer, Roberts, Lee, Narang, Matena,
  Zhou, Li and Liu}]{b7}
\bibinfo{author}{Raffel, C.}, \bibinfo{author}{Shazeer, N.},
  \bibinfo{author}{Roberts, A.}, \bibinfo{author}{Lee, K.},
  \bibinfo{author}{Narang, S.}, \bibinfo{author}{Matena, M.},
  \bibinfo{author}{Zhou, Y.}, \bibinfo{author}{Li, W.}, \bibinfo{author}{Liu,
  P.J.}, \bibinfo{year}{2020}.
\newblock \bibinfo{title}{Exploring the limits of transfer learning with a
  unified text-to-text transformer}.
\newblock \bibinfo{journal}{Journal of Machine Learning Research}
  \bibinfo{volume}{21}, \bibinfo{pages}{1--67}.
%Type = Inproceedings
\bibitem[{Reimers and Gurevych(2019)}]{b3}
\bibinfo{author}{Reimers, N.}, \bibinfo{author}{Gurevych, I.},
  \bibinfo{year}{2019}.
\newblock \bibinfo{title}{Sentence-bert: Sentence embeddings using siamese
  bert-networks}, in: \bibinfo{booktitle}{Proceedings of the 2019 Conference on
  Empirical Methods in Natural Language Processing and the 9th International
  Joint Conference on Natural Language Processing (EMNLP-IJCNLP)}, pp.
  \bibinfo{pages}{3982--3992}.
\newblock \DOIprefix\doi{10.18653/v1/D19-1410}.
%Type = Inproceedings
\bibitem[{Ren et~al.(2021)Ren, Qu, Liu, Zhao, She, Wu, Wang and Wen}]{b20}
\bibinfo{author}{Ren, R.}, \bibinfo{author}{Qu, Y.}, \bibinfo{author}{Liu, J.},
  \bibinfo{author}{Zhao, W.X.}, \bibinfo{author}{She, Q.}, \bibinfo{author}{Wu,
  H.}, \bibinfo{author}{Wang, H.}, \bibinfo{author}{Wen, J.R.},
  \bibinfo{year}{2021}.
\newblock \bibinfo{title}{Rocketqav2: A joint training method for dense passage
  retrieval and passage re-ranking}, in: \bibinfo{booktitle}{Proceedings of the
  2021 Conference on Empirical Methods in Natural Language Processing}, pp.
  \bibinfo{pages}{2825--2835}.
\newblock \DOIprefix\doi{10.18653/v1/2021.emnlp-main.224}.
%Type = Article
\bibitem[{Robertson and Zaragoza(2009)}]{b13}
\bibinfo{author}{Robertson, S.}, \bibinfo{author}{Zaragoza, H.},
  \bibinfo{year}{2009}.
\newblock \bibinfo{title}{The probabilistic relevance framework: Bm25 and
  beyond}.
\newblock \bibinfo{journal}{Found. Trends Inf. Retr.} \bibinfo{volume}{3},
  \bibinfo{pages}{333--389}.
\newblock \DOIprefix\doi{10.1561/1500000019}.
%Type = Inproceedings
\bibitem[{Ruder et~al.(2019)Ruder, Peters, Swayamdipta and Wolf}]{b14}
\bibinfo{author}{Ruder, S.}, \bibinfo{author}{Peters, M.E.},
  \bibinfo{author}{Swayamdipta, S.}, \bibinfo{author}{Wolf, T.},
  \bibinfo{year}{2019}.
\newblock \bibinfo{title}{Transfer learning in natural language processing},
  in: \bibinfo{booktitle}{Proceedings of the 2019 Conference of the North
  American Chapter of the Association for Computational Linguistics:
  Tutorials}, pp. \bibinfo{pages}{15--18}.
\newblock \DOIprefix\doi{10.18653/v1/N19-5004}.
%Type = Article
\bibitem[{Rudin(2019)}]{rudin2019stop}
\bibinfo{author}{Rudin, C.}, \bibinfo{year}{2019}.
\newblock \bibinfo{title}{Stop explaining black box machine learning models for
  high stakes decisions and use interpretable models instead}.
\newblock \bibinfo{journal}{Nature machine intelligence} \bibinfo{volume}{1},
  \bibinfo{pages}{206--215}.
%Type = Article
\bibitem[{Song et~al.(2023)Song, He, Yu, Fang, Cui and Lan}]{b25}
\bibinfo{author}{Song, C.}, \bibinfo{author}{He, H.}, \bibinfo{author}{Yu, H.},
  \bibinfo{author}{Fang, P.}, \bibinfo{author}{Cui, L.}, \bibinfo{author}{Lan,
  Z.}, \bibinfo{year}{2023}.
\newblock \bibinfo{title}{Uni-encoder: A fast and accurate response selection
  paradigm for generation-based dialogue systems}.
\newblock \bibinfo{journal}{Findings of the Association for Computational
  Linguistics: ACL} \bibinfo{volume}{2023}, \bibinfo{pages}{6231--6244}.
\newblock \DOIprefix\doi{10.18653/v1/2023.findings-acl.388}.
%Type = Inproceedings
\bibitem[{Tam et~al.(2023a)Tam, Liu, Ji, Xue, Liu, Li, Dong and
  Tang}]{tam2023parameter}
\bibinfo{author}{Tam, W.}, \bibinfo{author}{Liu, X.}, \bibinfo{author}{Ji, K.},
  \bibinfo{author}{Xue, L.}, \bibinfo{author}{Liu, J.}, \bibinfo{author}{Li,
  T.}, \bibinfo{author}{Dong, Y.}, \bibinfo{author}{Tang, J.},
  \bibinfo{year}{2023}a.
\newblock \bibinfo{title}{Parameter-efficient prompt tuning makes generalized
  and calibrated neural text retrievers}, in: \bibinfo{booktitle}{Findings of
  the Association for Computational Linguistics: EMNLP 2023}, pp.
  \bibinfo{pages}{13117--13130}.
%Type = Article
\bibitem[{Tam et~al.(2023b)Tam, Liu, Ji, Xue, Zhang, Dong, Liu, Hu and
  Tang}]{b26}
\bibinfo{author}{Tam, W.L.}, \bibinfo{author}{Liu, X.}, \bibinfo{author}{Ji,
  K.}, \bibinfo{author}{Xue, L.}, \bibinfo{author}{Zhang, X.},
  \bibinfo{author}{Dong, Y.}, \bibinfo{author}{Liu, J.}, \bibinfo{author}{Hu,
  M.}, \bibinfo{author}{Tang, J.}, \bibinfo{year}{2023}b.
\newblock \bibinfo{title}{Parameter-efficient prompt tuning makes generalized
  and calibrated neural text retrievers}.
\newblock \bibinfo{journal}{Findings of the Association for Computational
  Linguistics: EMNLP} \bibinfo{volume}{2023}, \bibinfo{pages}{13117--13130}.
\newblock \DOIprefix\doi{10.18653/v1/2023.findings-emnlp.87}.
%Type = Inproceedings
\bibitem[{Wallat et~al.(2023)Wallat, Beringer, Anand and
  Anand}]{wallat2023probing}
\bibinfo{author}{Wallat, J.}, \bibinfo{author}{Beringer, F.},
  \bibinfo{author}{Anand, A.}, \bibinfo{author}{Anand, A.},
  \bibinfo{year}{2023}.
\newblock \bibinfo{title}{Probing bert for ranking abilities}, in:
  \bibinfo{booktitle}{European Conference on Information Retrieval},
  \bibinfo{organization}{Springer}. pp. \bibinfo{pages}{255--273}.
%Type = Article
\bibitem[{Wang et~al.(2023)Wang, Wang, Wang, Naidu, Bergen and
  Paturi}]{wang2023scientific}
\bibinfo{author}{Wang, J.A.}, \bibinfo{author}{Wang, K.},
  \bibinfo{author}{Wang, X.}, \bibinfo{author}{Naidu, P.},
  \bibinfo{author}{Bergen, L.}, \bibinfo{author}{Paturi, R.},
  \bibinfo{year}{2023}.
\newblock \bibinfo{title}{Scientific document retrieval using multi-level
  aspect-based queries}.
\newblock \bibinfo{journal}{Advances in Neural Information Processing Systems}
  \bibinfo{volume}{36}, \bibinfo{pages}{38404--38419}.
%Type = Article
\bibitem[{Wang et~al.(2024)Wang, Chen and Verberne}]{wang2024quids}
\bibinfo{author}{Wang, Y.}, \bibinfo{author}{Chen, X.},
  \bibinfo{author}{Verberne, S.}, \bibinfo{year}{2024}.
\newblock \bibinfo{title}{Quids: Query intent generation via dual space
  modeling}.
\newblock \bibinfo{journal}{arXiv preprint arXiv:2410.12400} .
%Type = Article
\bibitem[{Yang et~al.(2024)Yang, Yang, Zhang, Hui, Zheng, Yu, Li, Liu, Huang,
  Wei, Lin, Yang, Tu, Zhang, Yang, Yang, Zhou, Lin, Dang, Lu, Bao, Yang, Yu,
  Li, Xue, Zhang, Zhu, Men, Lin, Li, Xia, Ren, Ren, Fan, Su, Zhang, Wan, Liu,
  Cui, Zhang and Qiu}]{qwen2.5}
\bibinfo{author}{Yang, A.}, \bibinfo{author}{Yang, B.}, \bibinfo{author}{Zhang,
  B.}, \bibinfo{author}{Hui, B.}, \bibinfo{author}{Zheng, B.},
  \bibinfo{author}{Yu, B.}, \bibinfo{author}{Li, C.}, \bibinfo{author}{Liu,
  D.}, \bibinfo{author}{Huang, F.}, \bibinfo{author}{Wei, H.},
  \bibinfo{author}{Lin, H.}, \bibinfo{author}{Yang, J.}, \bibinfo{author}{Tu,
  J.}, \bibinfo{author}{Zhang, J.}, \bibinfo{author}{Yang, J.},
  \bibinfo{author}{Yang, J.}, \bibinfo{author}{Zhou, J.}, \bibinfo{author}{Lin,
  J.}, \bibinfo{author}{Dang, K.}, \bibinfo{author}{Lu, K.},
  \bibinfo{author}{Bao, K.}, \bibinfo{author}{Yang, K.}, \bibinfo{author}{Yu,
  L.}, \bibinfo{author}{Li, M.}, \bibinfo{author}{Xue, M.},
  \bibinfo{author}{Zhang, P.}, \bibinfo{author}{Zhu, Q.}, \bibinfo{author}{Men,
  R.}, \bibinfo{author}{Lin, R.}, \bibinfo{author}{Li, T.},
  \bibinfo{author}{Xia, T.}, \bibinfo{author}{Ren, X.}, \bibinfo{author}{Ren,
  X.}, \bibinfo{author}{Fan, Y.}, \bibinfo{author}{Su, Y.},
  \bibinfo{author}{Zhang, Y.}, \bibinfo{author}{Wan, Y.}, \bibinfo{author}{Liu,
  Y.}, \bibinfo{author}{Cui, Z.}, \bibinfo{author}{Zhang, Z.},
  \bibinfo{author}{Qiu, Z.}, \bibinfo{year}{2024}.
\newblock \bibinfo{title}{Qwen2.5 technical report}.
\newblock \bibinfo{journal}{arXiv preprint arXiv:2412.15115} .
%Type = Article
\bibitem[{Yu et~al.(2024)Yu, Cohen, Lamba, Tetreault and
  Jaimes}]{yu2024explain}
\bibinfo{author}{Yu, P.}, \bibinfo{author}{Cohen, D.}, \bibinfo{author}{Lamba,
  H.}, \bibinfo{author}{Tetreault, J.}, \bibinfo{author}{Jaimes, A.},
  \bibinfo{year}{2024}.
\newblock \bibinfo{title}{Explain then rank: Scale calibration of neural
  rankers using natural language explanations from large language models}.
\newblock \bibinfo{journal}{arXiv e-prints} , \bibinfo{pages}{arXiv--2402}.
%Type = Inproceedings
\bibitem[{Yu et~al.(2022)Yu, Sun, Xu, Dong, Chen, Xu and
  Wen}]{yu2022explainable}
\bibinfo{author}{Yu, W.}, \bibinfo{author}{Sun, Z.}, \bibinfo{author}{Xu, J.},
  \bibinfo{author}{Dong, Z.}, \bibinfo{author}{Chen, X.}, \bibinfo{author}{Xu,
  H.}, \bibinfo{author}{Wen, J.R.}, \bibinfo{year}{2022}.
\newblock \bibinfo{title}{Explainable legal case matching via inverse optimal
  transport-based rationale extraction}, in: \bibinfo{booktitle}{Proceedings of
  the 45th international ACM SIGIR conference on research and development in
  information retrieval}, pp. \bibinfo{pages}{657--668}.
%Type = Unpublished
\bibitem[{Zhang et~al.(2021a)Zhang, Gong, Shen, Lv, Duan and Chen}]{b21}
\bibinfo{author}{Zhang, H.}, \bibinfo{author}{Gong, Y.}, \bibinfo{author}{Shen,
  Y.}, \bibinfo{author}{Lv, J.}, \bibinfo{author}{Duan, N.},
  \bibinfo{author}{Chen, W.}, \bibinfo{year}{2021}a.
\newblock \bibinfo{title}{Adversarial retriever-ranker for dense text
  retrieval}.
\newblock \href{http://arxiv.org/abs/vol}{{\tt arXiv:vol}}.
  \bibinfo{note}{abs/2110.03611}.
%Type = Inproceedings
\bibitem[{Zhang et~al.(2020)Zhang, Guo, Fan, Lan and Cheng}]{zhang2020query}
\bibinfo{author}{Zhang, R.}, \bibinfo{author}{Guo, J.}, \bibinfo{author}{Fan,
  Y.}, \bibinfo{author}{Lan, Y.}, \bibinfo{author}{Cheng, X.},
  \bibinfo{year}{2020}.
\newblock \bibinfo{title}{Query understanding via intent description
  generation}, in: \bibinfo{booktitle}{Proceedings of the 29th ACM
  International Conference on Information \& Knowledge Management}, pp.
  \bibinfo{pages}{1823--1832}.
%Type = Unpublished
\bibitem[{Zhang et~al.(2022)Zhang, Long, Xu and Xie}]{b24}
\bibinfo{author}{Zhang, Y.}, \bibinfo{author}{Long, D.}, \bibinfo{author}{Xu,
  G.}, \bibinfo{author}{Xie, P.}, \bibinfo{year}{2022}.
\newblock \bibinfo{title}{Hlatr: enhance multi-stage text retrieval with hybrid
  list aware transformer reranking}.
\newblock \href{http://arxiv.org/abs/2205.10569}{{\tt arXiv:2205.10569}}.
  \bibinfo{note}{arXiv preprint}.
%Type = Inproceedings
\bibitem[{Zhang et~al.(2019)Zhang, Han, Liu, Jiang, Sun and Liu}]{b8}
\bibinfo{author}{Zhang, Z.}, \bibinfo{author}{Han, X.}, \bibinfo{author}{Liu,
  Z.}, \bibinfo{author}{Jiang, X.}, \bibinfo{author}{Sun, M.},
  \bibinfo{author}{Liu, Q.}, \bibinfo{year}{2019}.
\newblock \bibinfo{title}{Ernie: Enhanced language representation with
  informative entities}, in: \bibinfo{booktitle}{Proceedings of the 57th Annual
  Meeting of the Association for Computational Linguistics, Jul. 2019}, pp.
  \bibinfo{pages}{1441--1451}.
\newblock \DOIprefix\doi{10.18653/v1/P19-1139}.
%Type = Inproceedings
\bibitem[{Zhang et~al.(2021b)Zhang, Rudra and Anand}]{zhang2021explain}
\bibinfo{author}{Zhang, Z.}, \bibinfo{author}{Rudra, K.},
  \bibinfo{author}{Anand, A.}, \bibinfo{year}{2021}b.
\newblock \bibinfo{title}{Explain and predict, and then predict again}, in:
  \bibinfo{booktitle}{Proceedings of the 14th ACM international conference on
  web search and data mining}, pp. \bibinfo{pages}{418--426}.

\end{thebibliography}

% %% For authoryear reference style
% %% \bibitem[Author(year)]{label}
% %% Text of bibliographic item

% \bibitem[Lamport(1994)]{lamport94}
%   Leslie Lamport,
%   \textit{\LaTeX: a document preparation system},
%   Addison Wesley, Massachusetts,
%   2nd edition,
%   1994.

% \end{thebibliography}
\end{document}